\documentclass[aps,10pt,twocolumn,prd,showpacs,showkeys,preprintnumbers,superscriptaddress,nobibnotes,floatfix,longbibliography,notitlepage,nofootinbib]{revtex4-1}

\usepackage{afterpage}
\usepackage{placeins}
\usepackage{amsmath}
\usepackage{amsfonts}
\usepackage{bm}
\usepackage{amssymb}
\usepackage{mathrsfs}
\usepackage{cancel}
\usepackage{accents}
\usepackage{mciteplus,slashed}
\usepackage{amssymb,cancel,amsmath,relsize}
\usepackage{mathrsfs} 
\usepackage{dcolumn}
\usepackage{bm}
\usepackage[caption=false]{subfig} 
\usepackage{appendix}
\usepackage{feynmp-auto}
\usepackage[T1]{fontenc}	
\usepackage{csvsimple}
\usepackage{hyperref}
\usepackage[capitalise]{cleveref}
\usepackage{booktabs}
\usepackage{graphicx}
\usepackage{mathrsfs}
\usepackage{floatpag}
\usepackage[utf8]{inputenc}
\usepackage[dvipsnames]{xcolor}
\hypersetup{
    colorlinks,
    linkcolor={red!50!black},
    citecolor={blue!50!black},
    urlcolor={blue!80!black}
}
\usepackage[normalem]{ulem}
\usepackage{cleveref}
\usepackage{fancybox}
\usepackage{array}
\newcolumntype{P}[1]{>{\centering\arraybackslash}p{#1}}
\newcolumntype{M}[1]{>{\centering\arraybackslash}m{#1}}
\usepackage{tabularray}
\usepackage{titlesec}

\titleformat*{\paragraph}{\bfseries \itshape}
\titleformat*{\section}{\centering\bfseries }
\titleformat*{\subsection}{\centering\bfseries }

\graphicspath{{./figures/}}

\begin{document}

\title{\Large Cosmological Consequences of Domain Walls\\ Biased by Quantum Gravity}

\author{Yann Gouttenoire}
\affiliation{School of Physics and Astronomy, Tel-Aviv University, Tel-Aviv 69978, Israel}
\affiliation{PRISMA+ Cluster of Excellence $\&$ MITP, Johannes Gutenberg University, 55099 Mainz, Germany}

\author{Stephen F. King}
\affiliation{School of Physics and Astronomy, University of Southampton, Southampton SO17 1BJ, United Kingdom}

\author{Rishav Roshan}
\affiliation{School of Physics and Astronomy, University of Southampton, Southampton SO17 1BJ, United Kingdom}

\author{Xin Wang}
\affiliation{School of Physics and Astronomy, University of Southampton, Southampton SO17 1BJ, United Kingdom}

\author{Graham White}
\affiliation{School of Physics and Astronomy, University of Southampton, Southampton SO17 1BJ, United Kingdom}

\author{Masahito Yamazaki}
\affiliation{Department of Physics, University of Tokyo, Hongo, Tokyo 113-0033, Japan}
\affiliation{Trans-Scale Quantum Science Institute,  University of Tokyo, Tokyo 113-0033, Japan}
\affiliation{Kavli IPMU (WPI), UTIAS, University of Tokyo, Chiba 277-8583, Japan}
\affiliation{Center for Data-Driven Discovery (CD3), Kavli IPMU, University of Tokyo, Chiba 277-8583, Japan}

\preprint{MITP-25-014}

\begin{abstract}

One of the simplest standard model extensions leading to a domain wall network is a real scalar $S$ with a $\mathcal{Z}_2$ symmetry spontaneously broken during universe evolution. Motivated by the swampland program, we explore the possibility that quantum gravity effects are responsible for violation of the discrete symmetry, triggering the annihilation of the domain wall network. We explore the resulting cosmological implications in terms of dark radiation, dark matter, gravitational waves, primordial black holes, and wormholes connected to baby universes.
\end{abstract}

\maketitle

\tableofcontents

\section{Introduction}
    \label{sec.introduction}

Despite advancements in quantum gravity (QG) and string theory over the past decades, it has proved challenging to find experimental tests of such high-scale physics. One recent development is the swampland program \cite{Vafa:2005ui,Ooguri:2006in}, which among other things suggests that exact global symmetries should be broken \cite{Banks:1988yz,Banks:2010zn,Harlow:2018tng,Fichet:2019ugl,Daus:2020vtf,King:2023ztb,Dvali:2018txx,Dvali:2017eba} in QG. Discrete global symmetries, such as $\mathbb{Z}_2$, arise in many theories beyond the Standard Model (SM), including dark matter (DM) and neutrino mass models (See, e.g. Refs.~\cite{ Farrar:1978xj, McDonald:1993ex,Burgess:2000yq, Altarelli:2010gt, Ishimori:2010au, King:2013eh}). If these symmetries are spontaneously broken during cosmological evolution, they result in the formation of a network of domain walls (DWs) \cite{Vilenkin:2000jqa}. During the matter- or radiation-dominated era, the fraction of the DW energy density $\rho_{\rm DW}$ relative to the radiation energy density $\rho_{\rm rad}$ increases linearly with time. This leads after a time ${\cal O}(M_{\rm pl}^2/\sigma)$ 
(where $M_{\rm pl}\simeq 2.4 \times 10^{18}~\rm GeV$ is the reduced Planck mass and $\sigma$ is the DW surface tension)
to a desolate universe dominated by DWs \cite{Zeldovich:1974uw},  in conflict with cosmological observations. On the other hand, a viable cosmological evolution can be achieved if the discrete symmetry is slightly broken, leading to a biased DW network that annihilates before it can dominate the energy density of the universe~\cite{Zeldovich:1974uw,Kibble:1976sj,Vilenkin:1981zs,Gelmini:1988sf,Larsson:1996sp}.
It is intriguing to consider that such symmetry breaking may arise from QG effects, as motivated by the swampland program \cite{Banks:1988yz,Banks:2010zn,Harlow:2018tng,Fichet:2019ugl,Daus:2020vtf,King:2023ztb}, thereby establishing a scenario where QG has direct cosmological implications.
QG effects responsible for the instabilities of both DWs and DM were explored in~\cite{King:2023ayw,King:2023ztb,Borah:2024kfn}. The model in \cite{King:2023ayw} considered two singlet scalar fields and two $\mathbb{Z}_2$ symmetries, one being responsible for DM stability, and the other spontaneously broken and responsible for DWs. Both $\mathbb{Z}_2$ symmetries were assumed to be explicitly broken by QG effective operators. The discussion was extended to the case of the fermionic DM in Refs.~\cite{King:2023ztb,Borah:2024kfn}. 

\begin{figure*}[t!]
     \centering
    \includegraphics[width=0.8\linewidth]{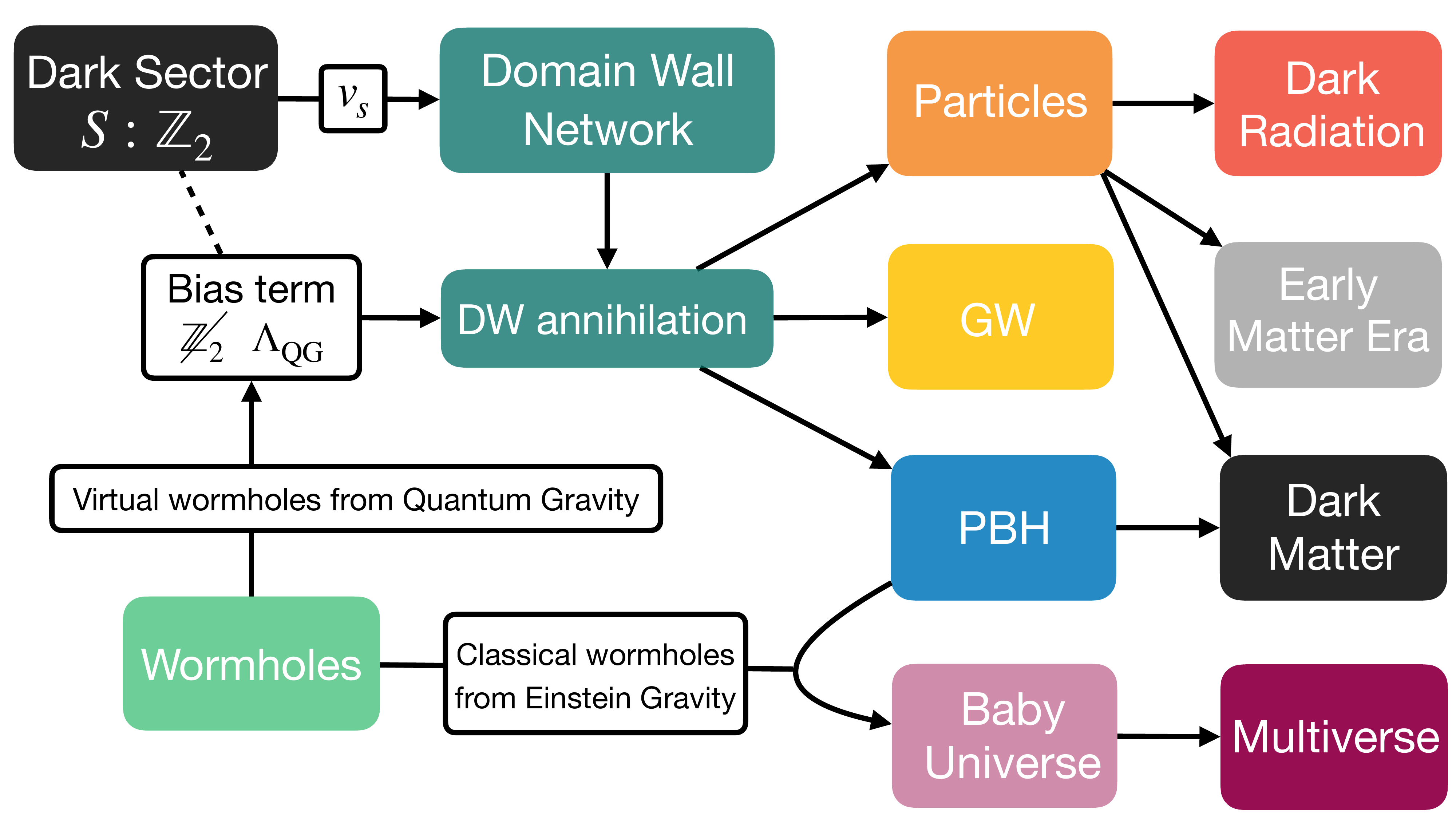}  
     \caption{A schematic illustration of the main ideas explored in this paper. A global ${\cal Z}_2$ symmetry in the dark sector is broken by QG effects leading to a biased DW network, which annihilates and generates Particles, GWs, or PBHs. Relativistic particles can contribute as Dark Radiation to $\Delta N_{\rm eff}$ (Sec.~\ref{sec:particleprod}). Non-relativistic particles can contribute to DM (Sec.~\ref{sec:darkmatter}) if stable or to an Early Matter Era if unstable and sufficiently abundant (Sec.~\ref{sec:matter_era}). GWs can be tested in current and future interferometers (Sec.~\ref{sec:GW formation}). PBHs arise from super-horizon DWs that annihilate sufficiently later than the others, allowing them to grow large enough to be contained within their Schwarzschild radius (Sec.~\ref{sec:PBH formation}). Wormholes enter this picture in two distinct ways. First, virtual wormholes --- arising in the path integral in QG --- can serve as non-perturbative sources of global symmetry breaking~\cite{Giddings:1987cg,Maldacena:2004rf,Arkani-Hamed:2007cpn,Hebecker:2018ofv,VanRiet:2020pcn,Daus:2020vtf}. Second, classical wormholes can form within Einstein gravity when DWs inflate into ``baby universes'' (Sec.~\ref{sec:wormhole_formation}). This occurs if the DWs grow larger than the Hubble horizon of the universe dominated by the vacuum bias $V_{\rm bias}$, experienced by observers located inside the DW. In this scenario, the DW induces a classical wormhole geometry connecting the PBH to the baby universe. Being eternally-inflating, those baby universe leads to the existence of a multiverse.  In short, we are left with the intriguing sequence of events where virtual wormholes are potentially responsible for the discrete symmetry breaking that sets the stage for classical wormholes connecting our universe to a multiverse.  }
     \label{fig:schematic-2} 
 \end{figure*}
In this paper, we consider a scenario even more minimal than those in \cite{King:2023ayw,King:2023ztb,Borah:2024kfn}, with a \emph{single} $\mathbb{Z}_2$ symmetry which is spontaneously broken by a scalar field that acquires a vacuum expectation value (VEV) $v_s$ in the early universe. The $\mathbb{Z}_2$-symmetry $v_s\leftrightarrow -v_s$ is broken explicitly by a bias term $V_{\rm bias}$ of QG origin. The bias pressure drives the DW network to suddenly annihilate when it overcomes the forces due to the surface tension $\sigma$ after a time $t_{\rm ann}\sim \sigma/V_{\rm bias}$. The smallness of the QG bias term $V_{\rm bias}$ results in a long lifetime of the DW network, and a large energy fraction $\rho_{\rm DW}/\rho_{\rm rad}\sim \sigma^2/(M_{\rm pl}^2V_{\rm bias})$ at the epoch of annihilation, leading to efficient production of particles~\cite{Vachaspati:1984yi,Watkins:1991zt,Kolb:1996jr,Falkowski:2012fb,Shakya:2023kjf,Mansour:2023fwj,DEramo:2024lsk,Daido:2015gqa,Mariotti:2024eoh,Vanvlasselaer:2024vmi,Azzola:2024pzq}, gravitational waves (GWs) \cite{Vilenkin:1981zs,Chang:1998tb,Gleiser:1998na,Preskill:1991kd,Hiramatsu:2010yz,Kawasaki:2011vv,Saikawa:2017hiv,Roshan:2024qnv}, primordial black holes (PBHs) \cite{Ferrer:2018uiu,Gelmini:2023ngs,Gelmini:2022nim,Gouttenoire:2023gbn,Gouttenoire:2023ftk,Ferreira:2024eru,Lu:2024ngi}, and wormholes connected to baby universe~\cite{Gouttenoire:2023gbn}, see Fig.~\ref{fig:schematic-2}. The particles are produced from reheating after DW network annihilation. They can contribute to the number of effective neutrino flavors $N_{\rm eff}$~\cite{Dvorkin:2022jyg}, Dark Matter (DM)~\cite{DEramo:2024lsk} and if sufficiently abundant, they can even generate an early matter era, e.g.~\cite{Weinberg:1982zq}. The mechanism of PBH formation from the annihilation of a biased DW network studied here~\cite{Ferrer:2018uiu,Gelmini:2023ngs,Gelmini:2022nim,Gouttenoire:2023gbn,Gouttenoire:2023ftk,Ferreira:2024eru,Lu:2024ngi} differs from the one predicted in DW networks bounded by cosmic strings \cite{Ge:2019ihf,Ge:2023rrq,Dunsky:2024zdo}; spherical domains produced during inflation~\cite{Garriga:2015fdk,Deng:2016vzb,Deng:2017uwc,Deng:2020mds,Kusenko:2020pcg,Maeso:2021xvl,Escriva:2023uko,He:2023yvl,Huang:2023mwy,Kitajima:2020kig,Kasai:2023ofh,Kasai:2023qic}; or false vacuum regions formed during first-order phase transition~\cite{Kodama:1982sf,Hsu:1990fg,Liu:2021svg,Hashino:2021qoq,Kawana:2022olo,Lewicki:2023ioy,Gouttenoire:2023naa,Baldes:2023rqv,Gouttenoire:2023bqy,Salvio:2023ynn,Gouttenoire:2023pxh,Jinno:2023vnr,Flores:2024lng,Lewicki:2024ghw,Lewicki:2024sfw,Cai:2024nln,Ai:2024cka,Arteaga:2024vde,Banerjee:2024cwv,Hashino:2025fse}.  The mechanism of baby universe production developed here~\cite{Gouttenoire:2023gbn} is classical and takes place in the radiation-dominated phase.\footnote{The origin of the scalar potential may, however, require non-classical dynamics, see the discussion at the end of Sec.~\ref{sec:wormhole_formation}.} It differs from the usual baby universe creation from quantum or thermal tunneling during inflation~\cite{Goncharov:1987ir,Linde:1990ta,Linde:1991sk,Garriga:2015fdk,Deng:2016vzb}.
  
The paper is organized as follows. In Sec.~\ref{sec:simplified}, we introduce a simplified model involving a real scalar field charged under a $\mathbb{Z}_2$-symmetry, responsible for the formation of a DW network, and discuss sources of explicit symmetry breaking arising from QG. 
We then review the annihilation of DWs and the resulting generation of dark radiation in Sec.~\ref{sec:particleprod}, early matter era in Sec.~\ref{sec:matter_era}, DM in Sec.~\ref{sec:darkmatter}, GWs in  Sec.~\ref{sec:GW formation}, PBHs in Sec.~\ref{sec:PBH formation}, and baby universes in Sec.~\ref{sec:wormhole_formation}. 
Finally, in Sec.~\ref{sec:conclusion}, we present our conclusions.

\section{A simple discrete symmetry broken by QG} \label{sec:simplified}
\subsection{A singlet scalar charged under \texorpdfstring{$\mathbb{Z}_2$}{Z(2)}}

\paragraph*{The Lagrangian.}
In this work, we consider a minimal extension of the SM by adding a real scalar field $S$. This field $S$ is neutral under the SM gauge symmetry but is charged under a new discrete $\mathbb{Z}_2$ symmetry which does not affect SM particles. Since string theory predicts ${\cal O}(100)$ different moduli fields \cite{Witten:1984dg,Svrcek:2006yi,Arvanitaki:2009fg,Cicoli:2012sz,Demirtas:2018akl,Visinelli:2018utg,Carta:2020ohw,Mehta:2021pwf,Cicoli:2021gss,Dimastrogiovanni:2023juq}, it is natural to expect that some of them can play the role of the $S$ field. Such an extension can also lead to a strong first-order electroweak (EW) phase transition as needed for EW baryogenesis~\cite{McDonald:1993ey,Espinosa:1993bs,Choi:1993cv,Blasi:2022woz,Blasi:2023rqi,Agrawal:2023cgp}.

The most general renormalizable scalar potential involving the SM Higgs doublet $H$ and the $\mathbb{Z}_2$-odd real scalar singlet $S$ at tree level can be expressed as:
\begin{eqnarray}
    V&=& \frac{\mu ^2}{2}  H^\dagger H+ \frac{\lambda_h}{4} (H^\dagger H)^2 + \frac{\lambda _{hs}}{4} H^\dagger H S^2  \nonumber 
    \\ 
    && \hspace{0cm}+\frac{\lambda_s}{4} (S^2-v_s^2)^2  \; .
    \label{eq:lagrangian_tree}
\end{eqnarray}
The scalar potential should be bounded from below to make the electroweak vacuum stable, which leads to the following constraints~\cite{Kannike:2012pe,Chakrabortty:2013mha}:
\begin{eqnarray}
    \lambda_h,\lambda_s\geq 0 \; ,\qquad \lambda_{hs}+2\sqrt{\lambda_h\lambda_s}\geq 0 \;.
\end{eqnarray}

After the EW symmetry breaking, the CP-even component of the Higgs field acquires its VEV $v_h$. We introduce the scalar excitations $h$ and $s$ of the Higgs doublet $H = (0,v_h+h)$ and dark singlet $S=v_s+s$ around their VEV, respectively. The interactions between scalar excitations $h$ and $s$ is controlled by the mixing angle $\theta_{hs}$, defined as the rotation angle between the gauge eigenstates and the mass eigenstates~\cite{Bhattacharya:2019tqq,Gouttenoire:2023pxh}:
\begin{equation}
\label{eq:theta_hs}
    \tan{(2\theta_{hs})} = \frac{\lambda_{hs}v_hv_s}{\lambda_s v_s^2 - \lambda_h v_h^2} \; .
\end{equation}
We work in the limit $v_s \gg v_h$ and $\lambda^{}_{hs} \ll 1$, where $\theta_{hs}$ is negligibly small
unless there exists significant fine-tuning such that $\lambda_s v_s^2 - \lambda_h v^2_h\sim 0$.
 
\paragraph*{DW network formation.}
At high temperature, the potential receives finite-temperature corrections which maintain the field at the $\mathbb{Z}_2$-symmetric value $\left<S\right>=0$ \cite{Kirzhnits:1972ut,Dolan:1973qd, Weinberg:1974hy}. More specifically, the thermal mass of the singlet scalar around the origin $S=0$ \cite{Weinberg:1974hy} reads:
\begin{equation}
    \mu_{\rm s,th}^{2}\simeq \left(\frac{\lambda_{hs}}{6}+\frac{\lambda_s}{4}\right)T^2\simeq \frac{1}{4}\lambda_s T^2 \; ,
\end{equation}
whose magnitude is initially larger than that of the negative zero-temperature mass $\mu_{s,T=0}^2 \equiv \partial^2V/\partial S^2\big|_{S=0}= -\lambda_s v_s^2$.
When the temperature drops below 
\begin{equation}
\label{eq:T_form}
    T_{\rm form}\simeq  \frac{2v_s}{\sqrt{1+2\lambda_{hs}/3\lambda_s}}\simeq 2v_s \; ,
\end{equation}
 the zero-temperature mass dominates over the thermal mass, and the universe undergoes a phase transition during which the scalar $S$ develops a non-zero VEV $\langle S \rangle = \pm v_s$, distinct for each causal patch. This leads to the formation of a network of DWs~\cite{Zeldovich:1974uw,Kibble:1976sj}, defined as boundaries between the domains associated with different field values $+ v_s$ and $-v_s$. 

In order for the DW network to form successfully, the temperature $T_{\rm form}$ cannot be higher than the maximal temperature $T_{\rm max}$ of the universe, $T_{\rm form} \lesssim T_{\rm max}$. 
Assuming instantaneous reheating after inflation, the latter is given by:
\begin{equation}
\label{eq:Tmax_uni}
    T_{\rm max}\simeq \left(\frac{3M_{\rm pl}^2H_{\rm end}^2}{\pi^2g_{\star}/30}\right)^{\!\!1/4} \; ,
\end{equation}
where $H_{\rm end}$ is the Hubble scale at the end of inflation and $g_\ast$ is the number of relativistic degrees of freedom. The Hubble scale during inflation is directly related to the tensor power spectrum $\Delta_t^2(k)$ via~\cite{Baumann:2022mni}:
\begin{equation}
    \Delta_t^2(k) = \frac{2}{\pi^2}\frac{H_k^2}{M_{\rm pl}^2} \; .
\end{equation}
$\Delta_t^2(k)$ is usually modeled as a power-law $\Delta_t^2(k)= A_t\left( k/k_{\star}\right)^{n_t}$
with $k_{\star}\equiv 0.05~\rm Mpc^{-1}$, $A_t$ being the amplitude, and $n_t$ denoting the spectral index.
The non-observation of primordial B modes in CMB anisotropies by BICEP/Keck~\cite{BICEP:2021xfz} provides the most stringent upper bound on the tensor-to-scalar ratio $r\equiv A_t/A_s\lesssim 0.036$, where $A_s\simeq 2.099\times 10^{-9}$ is the curvature power spectrum measured by Planck~\cite{Aghanim:2018eyx}, with both $r$ and $A_s$ being evaluated at the CMB scale $k_{\star}$.
The BICEP/Keck bound on $r$ \cite{BICEP:2021xfz} translates to an upper bound on the maximal reheating temperature of the universe:
\begin{equation}
\label{eq:T_max}
    T_{\rm max} \simeq 5.8\times 10^{15}~{\rm GeV}\!\left(\frac{r}{0.036} \right)^{\!\!1/4}\!\left(\frac{107.75}{g_{\star}} \right)^{\!\!1/4}\!\!\left(\frac{H_{\rm end}}{H_{\star}} \right)^{\!\!1/2}\!\!,
\end{equation}
where $H_{\star}$ is the inflation scale when the CMB scale $k_{\star}=0.05~\rm Mpc^{-1}$ exits the horizon. In order to obtain the last factor in the above equation, we have used the relation $\left({H_{\rm end}/H_{\star}} \right)^{\!1/2}= \left({k_{\rm end}}/{k_{\star}} \right)^{\!n_t/2}$. We assume $H_{\rm end}\simeq H_{\star} $ such that the last factor is simply unity. 

Furthermore, DW formation requires that $V_{\rm bias}$ at DW formation must be smaller than the barrier $V_b$ separating distinct vacua~\cite{Gelmini:1988sf}. We have checked that this constraint is safely satisfied throughout this work.

\subsection{QG effective operators}

The discrete $\mathbb{Z}_2$ symmetry in the Lagrangian in Eq.~\eqref{eq:lagrangian_tree} is expected to be explicitly violated in QG theories \cite{Banks:1988yz,Banks:2010zn,Harlow:2018tng}.\footnote{There exists a much stronger conjecture that such a discrete symmetry breaking is impossible in QG, see \cite{Dvali:2018txx}. While we will not assume this conjecture in this paper, our discussion provides an empirical test of the conjecture.
} We operate under the hypothesis, denoted as the Refined Swampland Global Symmetry Conjecture in \cite{King:2023ztb}, that the QG effective operators breaking the $\mathbb{Z}_2$ symmetry are of the lowest possible dimension, which in our case is dimension-five:
\begin{eqnarray}
    \Delta V  &=& \frac{1}{\Lambda_{\rm QG}} (\alpha_{1}  S^5 + \alpha_{2}  S^3 |H|^2 + \alpha_{3}  S |H|^4 ) \; ,
    \label{eq:bias}
\end{eqnarray}
where $|H|^2=H^\dagger H$ and we will henceforth set $\alpha_1 = 1$ by rescaling the value of $\Lambda_{\rm QG}$. These operators are the source of the QG bias term responsible for the annihilation of the DW network.

In the literature, the dimensional scale $\Lambda_{\rm QG}$ associated with the higher-dimensional operators that break the global symmetry is often assumed to be the reduced Planck mass $M_{\rm pl}$~\cite{Addazi:2021xuf}.
While this is natural when the symmetry breaking is caused by perturbative gravity effects, 
the size of the symmetry-breaking effects may be further suppressed by non-perturbative effects, leading to an effective breaking scale many orders of magnitude higher than $M^{}_{\rm pl}$:
\begin{align} 
    \label{Lambda_QG}
    \Lambda_{\rm QG} = M_{\rm pl}\, e^{\mathcal{S}} \gg M_{\rm pl} \; ,
\end{align} 
where $\mathcal{S}$ represents the size of the action of the non-perturbative instanton. 

Wormholes \cite{Giddings:1987cg} are among the potential sources of non-perturbative global symmetry breaking. Wormholes are sources of apparent paradoxes in holography \cite{Maldacena:2004rf,Arkani-Hamed:2007cpn}
and their relevance in the path integral of quantum gravity has been under active discussion (see e.g.\ \cite{Hebecker:2018ofv,VanRiet:2020pcn} for an overview). Wormholes have interesting phenomenological consequences, as reviewed, e.g.\ in Ref.~\cite{Hebecker:2018ofv}, and 
the relevance of their non-perturbative contributions is suggested \cite{Daus:2020vtf} by the axionic version of the weak gravity conjecture \cite{Arkani-Hamed:2006emk}. 

\subsection{DW networks with QG bias}

\paragraph*{DW scaling regime.}
The dynamical evolution of the DW network is determined by the interplay between two effects: the Hubble expansion, which tends to increase its energy fraction with respect to the radiation background, and its surface tension $\sigma$, which tends to convert the smallest structures into scalar radiation and gravitational waves. These complex dynamics can be well described by a scaling law in which the correlation length of the network is set by the cosmic horizon $L\simeq t$ \cite{Press:1989yh}. The DW energy density $\rho_{\rm DW}$ can be determined by dimensional analysis \cite{Vilenkin:2000jqa,Saikawa:2017hiv}:
\begin{equation}
\label{eq:rho_DW_A}
    \rho_{\rm DW} = \frac{\sigma}{L} \; ,\qquad \textrm{with}~L\simeq  t/\mathcal{A} \; ,
\end{equation} 
where the area parameter has been calculated to be $\mathcal{A} = 0.8 \pm 0.1$ \cite{Hiramatsu:2013qaa} for the $\mathbb{Z}_2$-symmetric potential. The surface tension $\sigma$ is given by~\cite{Saikawa:2017hiv}:
\begin{equation}
    \label{eq:sigma}
    \sigma = \int_{-\infty}^{+\infty}dz\, T_{00} = \frac{4}{3} \sqrt{\frac{\lambda_s}{2}} v_s^3 \; ,
\end{equation}
with $T_{00} = (d\phi/dz)^2$ being the 00-component of the energy-momentum tensor. 
We have substituted the kink solution $\phi(z) = v_s \tanh(\sqrt{\lambda_s/2}\, v_s z)$ for the DW separating two degenerate vacua to obtain the final result in Eq.~(\ref{eq:sigma}).  The redshift of DWs in Eq.~\eqref{eq:rho_DW_A} is slower than that of the radiation background $\rho_{\rm rad} \simeq \pi^2 g_{\star} T^4/30 \propto 1/t^2$, hence DWs may dominate the energy density of the universe when $\rho_{\rm DW}\simeq 3M_{\rm pl}^2H^2$. This is the famous DW problem \cite{Zeldovich:1974uw}. 

\paragraph*{Biased DW networks.}
The simplest solution to the DW problem is to introduce a bias term that lifts the degeneracy between two vacua, making the DW network unstable \cite{Zeldovich:1974uw, Kibble:1976sj,Vilenkin:1981zs, Gelmini:1988sf, Larsson:1996sp}. Although bias terms are typically added manually, in this paper, we consider global symmetry breaking operators that are naturally generated in QG. The dimension-five operators in Eq.~(\ref{eq:bias}) contribute to the effective potential in the following form:
\begin{equation}
    V_{\rm bias} \simeq \frac{1}{\Lambda_{\rm QG}} \left( v_s^5 + \alpha_2 \frac{v_s^3 v_h^2}{2} + \alpha_3 \frac{v_s v_h^4}{4} \right) \; .
    \label{eq:model-bias}
\end{equation}
For simplicity, we assume that the first term in Eq.~(\ref{eq:model-bias}) dominates over the other two, giving us:
\begin{equation}
    V_{\rm bias} \simeq \frac{v_s^5}{\Lambda_{\rm QG}} \; .
    \label{eq:model-bias1}
\end{equation}
This assumption is justified, e.g.\ when we consider $ v_s \gg v_h$ and the dimensionless coefficients $\alpha_i$ in Eq.~\eqref{eq:bias} are of the same order.\footnote{Here we follow the motivation in \cite{King:2023ayw}: Suppose that we are interested in a class of string-theory compactifications. Even if we do not know all the details of compactifications, with the help of the swampland constraint, it is plausible to consider a constraint $\alpha_i\lesssim c$ for some constant $c$. Then the choice $\alpha_i\sim c$ would invoke global symmetries of the best quality at low energy, which is not forbidden by the current knowledge of the swampland. Of course, this is {\it not} the argument for a stronger statement that $\alpha_i$ should always be of the same order.}

The surface tension $\sigma$ in Eq.~\eqref{eq:sigma} as well as the size of the bias $V_{\rm bias}$ in Eq.~\eqref{eq:model-bias1} will be
crucial parameters determining the fate of the DWs. Note that the 
two DW parameters $\sigma$ and $V^{}_{\rm bias}$ depend only on two combinations of the three model parameters $v_s, \lambda_s$ and 
$\Lambda_{\rm QG}$, which we can choose to be:
\begin{equation}
    \label{eq:combination}
    \lambda_s^{1/6} v_s \;, \quad \lambda_s^{5/6} \Lambda_{\rm QG} \;.
\end{equation}
This means that the effect of $\lambda_s$ can be absorbed into the rescaling of $v_s$ and $\Lambda_{\rm QG}$.

\paragraph*{DW energy fraction.}
The bias term $V_{\rm bias}$ in Eq.~\eqref{eq:model-bias1} sources a pressure on the wall. When such a pressure is larger than the pressure $C_d\, \sigma / L$ arising from the surface tension $\sigma$, the DW network starts to annihilate. This occurs after the time:
\begin{equation}
    \label{eq:t_ann}
    t_{\rm ann} \simeq  C_d \,\mathcal{A}\,\frac{\sigma}{V_{\rm bias}} \;,
\end{equation}
where $C_d\sim 3$ according to lattice simulations \cite{Kawasaki:2014sqa}. To be cosmologically viable, the DW network must annihilate before dominating the universe, i.e. 
\begin{equation}
    t_{\rm ann} \lesssim t_{\rm dom} \; ,
\end{equation}
where we define $t_{\rm dom }$ as the time when false vacuum regions are dominated by the energy density of the biased DW network
\begin{equation}
\label{eq:t_dom_def}
\mathcal{A}\sigma/t_{\rm dom}+V_{\rm bias} \simeq \rho_{\rm rad}(t_{\rm dom}) \; .
\end{equation}
For simplicity, we approximate $\rho_{\rm rad}= \rho_{\rm ann}(a_{\rm ann}/a)^4 \simeq 3M_{\rm pl}^2/(2t)^2$ as in the case of pure-radiation domination, even though the universe contains DWs and vacuum energy components. We obtain
\begin{equation}
    \label{eq:t_dom}
     t_{\rm dom }~ \simeq~ \frac{\sqrt{3C_d^2M_{\rm pl}^2/V_{\rm bias}+t_{\rm ann}^2}-t_{\rm ann}}{2C_d} ~\simeq~ \frac{\sqrt{3}M_{\rm pl}}{2\sqrt{V_{\rm bias}}} \; ,
\end{equation}
where the last term neglects higher-order corrections in $\mathcal{O}(t_{\rm ann}/t_{\rm dom})$. The total energy fraction of the biased DW network reads 
\begin{align}
\label{eq:alpha_ann_DW_V}
    \alpha^{\rm DW+V}_{\rm ann} &\simeq \frac{\rho_{\rm DW}+\rho_{\rm V}}{3M_{\rm pl}^2H^2} \Big|_{t=t_{\rm ann}}\simeq (1+C_d/2)\alpha_{\rm ann}^{\rm DW} \; ,
\end{align}
where $\rho_{\rm DW}\simeq \mathcal{A}\sigma/t$ is the average DW energy density (surface contribution) and $\rho_{\rm V}\simeq V_{\rm bias}/2$ is the average vacuum bias energy density (bulk contribution).\footnote{Note that the factor $1/2$ in $\rho_{\rm V}\simeq V_{\rm bias}/2$ did not appear in Eq.~\eqref{eq:t_dom_def} since we defined $t_{\rm dom}$ as the time of domination from the point of view of an observer inside the false vacuum. We chose this definition to match the minimal time $t_{\rm PBH}$ of PBH formation in Eq.~\eqref{eq:t_PBH_t_dom} found after solving the equation of motion of a thin shell in General Relativity.} We have introduced in Eq.~(\ref{eq:alpha_ann_DW_V}) the DW surface energy fraction $\alpha_{\rm ann}^{\rm DW}$ at the time $t_{\rm ann}$ of DW annihilation:
\begin{align}
    \label{eq:alpha_DW}
     \alpha_{\rm ann}^{\rm DW} \equiv \frac{\rho_{\rm DW}}{3M_{\rm pl}^2H^2} \Big|_{t=t_{\rm ann}}
     \simeq C_d^{-1}\left( \frac{t_{\rm ann}}{t_{\rm dom }} \right)^2 \; ,
\end{align}
where we have used the approximated relation $H\simeq 1/2t$.\footnote{If instead we consider the full expression 
$H^2 = (\rho_{\rm rad}+\rho_{\rm DW}+\rho_{\rm V})/3M_{\rm pl}^2 =[ 1+\alpha_{\rm ann}(t/t_{\rm ann})+C_d\alpha_{\rm ann}(t/t_{\rm ann})^{2}]/(8t^2)$, then Eq.~\eqref{eq:alpha_DW} becomes $ \alpha_{\rm ann}^{\rm DW} \simeq \tilde{\alpha}_{\rm ann}^{\rm DW} (1+\sqrt{\tilde{\alpha}_{\rm ann}^{\rm DW}/C_d}+(2+C_d)\tilde{\alpha}_{\rm ann}^{\rm DW}/2) $ with $\tilde{\alpha}_{\rm ann}^{\rm DW}\equiv (t_{\rm ann}/t_{\rm dom })^2/C_d$ where we assumed $t_{\rm dom}$ given by the first equality in Eq.~\eqref{eq:t_dom}. }
Since it is the quantity $\alpha_{\rm ann}^{\rm DW}$ which enters the GW spectrum from DW annihilation, see Eq.~\eqref{eq:GW_DW_ann} below, in what follows we express most of the quantities in terms of $\alpha_{\rm ann}\equiv \alpha_{\rm ann}^{\rm DW}$ defined in Eq.~\eqref{eq:alpha_DW}. Nevertheless, $\alpha_{\rm ann}^{\rm DW+V}$ is relevant when we discuss the particle production in Eq.~\eqref{eq:rho_DR_T}. 
Using Eqs.~\eqref{eq:sigma} and \eqref{eq:model-bias1}, we can express $\alpha_{\rm ann}$ in terms of the parameters of the particle physics + QG model studied in this paper:
\begin{equation}
\label{eq:alpha_DW_more}
    \alpha_{\rm ann} \equiv \alpha_{\rm ann}^{\rm DW} \simeq \frac{32C_d{\cal A}^2}{27 M_{\rm pl}^2}
    (\lambda_s^{5/6}\Lambda_{\rm QG})(\lambda_s^{1/6} v_s) \; .
\end{equation}

\section{Consequences of long-lived DW networks}
    \label{sec:origin}

Previously, we introduced a $\mathbb{Z}_2$ singlet scalar. We have shown that it leads to a DW network with an energy density fraction growing linearly with time $\rho_{\rm DW}/\rho_{\rm rad}\propto t$ and annihilating at time $t_{\rm ann}$. The $\mathbb{Z}_2$-breaking correction from QG leads the DWs to annihilate with the energy density fraction $\alpha_{\rm ann}$ given in Eq.~\eqref{eq:alpha_DW}.  We now discuss the consequences of DW annihilation occurring when $\alpha_{\rm ann}$ is large.

\subsection{Dark Radiation}
\label{sec:particleprod}
During annihilation, most of the DW energy is transferred to scalar particles $s$ and any particles coupled to it~\cite{Widrow:1989vj}.  If they behave as dark radiation (DR), they can contribute to $N_{\rm eff}$; alternatively, if they behave as non-relativistic matter, they can contribute to a matter era.
In the case where DWs annihilate into dark degrees of freedom, the DW annihilation produces DR with energy density: 
\begin{align}
    \rho_{\rm DR}(T) &= \alpha_{\rm ann}^{\rm DW+V} \rho_{\rm rad}(T_{\rm ann})(a(T_{\rm ann})/a(T))^4 \; ,
\end{align}
where the quantity $\alpha_{\rm ann}^{\rm DW+V}$ is defined in Eq.~\eqref{eq:alpha_ann_DW_V}. Using $\rho\propto g_{\star}T^4$ and $a\propto g_{\star,s}(T)^{-1}T^{-1}$ (with $g_{\star,s}$ being the effective number of relativistic degrees of freedom entering the entropy density), we obtain:\footnote{We recall that $g_\star(T_0)\equiv 2+(7/8)\cdot 2\cdot N_{\rm eff}\cdot (4/11)^{4/3} \simeq 3.38$, and $g_{\star,s}(T_0)\equiv 2+(7/8)\cdot 2\cdot N_{\rm eff}\cdot (4/11) \simeq 3.94$ when we assume $N_{\rm eff}\simeq 3.043$ \cite{Cielo:2023bqp}.}
\begin{align}
\label{eq:rho_DR_T}
    \rho_{\rm DR}(T)    &=  \alpha_{\rm ann}^{\rm DW+V}\rho_{\rm rad}(T)\left( \frac{g_{\star}(T_{\rm ann})}{g_{\star}(T)} \right)\!\left( \frac{g_{\star,s}(T)}{g_{\star,s}(T_{\rm ann})} \right)^{\!4/3} \; ,
\end{align}
which contributes to the number of neutrino flavors as follows:
\begin{equation}
\label{eq:Delta_Neff}
 \Delta N_{\rm eff}(T_0) = \frac{8}{7}\left( \frac{11}{4} \right)^{4/3}\!\!\left( \frac{\rho_{\rm DR}(T_0)}{\rho_\gamma(T_0)} \right) \; ,
\end{equation}
where $\rho_{\gamma}$ is the photon number density and $T_0$ is any temperature below neutrino decoupling temperature $T_0\lesssim \rm MeV$.
The parameter $\Delta N_{\rm eff}$ is constrained to $\Delta N_{\rm eff} \lesssim 0.34$ at the 95$\%$ confidence level ($2\sigma$) by the combined BAO and Planck data~\cite{Planck:2018vyg, Dvorkin:2022jyg}, with a slight potential improvement over the BBN data~\cite{Pitrou:2018cgg}. Plugging Eq.~\eqref{eq:rho_DR_T} into Eq.~\eqref{eq:Delta_Neff}, we obtain:
\begin{align}
\label{eq:Delta_Neff_bound}
    \Delta N_{\rm eff}     &=  \frac{8}{7}\left( \frac{11}{4} \right)^{\!4/3}\!\!\left( \frac{g_{\star}(T_{\rm ann})}{g_\gamma} \right)\!\left( \frac{g_{\star,s}(T_0)}{g_{\star,s}(T_{\rm ann})} \right)^{\!4/3}\!\!\!\!\alpha_{\rm ann}^{\rm DW+V}\; ,\notag
\end{align}
where $g_\gamma=2$ is the number of photon degrees of freedom.  Plugging numerical values, we obtain the following constraints on DWs annihilating into DR:
\begin{equation}
\label{eq:Delta_N_eff_app}
    \Delta N_{\rm eff} = 7.4\left( \frac{g_{\star}(T_{\rm ann})}{g_{\star}(T_0)} \right)\!\left( \frac{g_{\star,s}(T_0)}{g_{\star,s}(T_{\rm ann})} \right)^{\!4/3}\!\!\!\!\alpha_{\rm ann}^{\rm DW+V} \lesssim 0.34 \; .
\end{equation}
For DWs annihilating into DR, the previous bound applies for all values of $T_{\rm ann}$. Instead, for DW annihilation into the SM, the bound in Eq.~\eqref{eq:Delta_N_eff_app} applies only for DWs annihilating below the neutrino decoupling temperature $T_{\rm ann}\lesssim 1~\rm MeV$.  In Figs.~\ref{fig:DW_generic} and \ref{fig:BrokenZ2_QG_PBHs}, we show the BBN constraints assuming that DWs annihilate into SM or dark radiation.

\subsection{Early Matter Era}
\label{sec:matter_era}
We now examine the scenario in which DW annihilation products manifest as long-lived, non-relativistic particles. 
As indicated by the Lagrangian in Eq.~\eqref{eq:lagrangian_tree}, the decay width of the singlet scalar field $s$ is given by:
\begin{equation}
\label{eq:def_Gamma_s}
\Gamma_s = \Gamma_{s\to hh}+\sin^2(\theta_{hs})\Gamma_h(m_s).
\end{equation}
The first decay channel is the decay into SM Higgs bosons:
\begin{align}
    \Gamma_{s\to hh} &\simeq \frac{\lambda_{hs}^2v_s^2}{32\pi m_s} \; ,
\end{align}
where $m_s^2=2\lambda_sv_s^2$. The other decay channel is induced by the Higgs mixing whose angle is given in Eq.~\eqref{eq:theta_hs}, which for $\theta_{h\phi}\ll 1$, gives:
\begin{align}
\label{eq:decay_s_SM}
\Gamma_h(m_s)\sin^2(\theta_{hs})&\simeq \lambda_{hs}^2\Gamma_h(m_h) v_h^2v_s^2 
    \begin{cases}
m_s^{-4},\quad m_s\gg m_h,\\
m_h^{-4},\quad m_s\ll  m_h,
    \end{cases}\notag
\end{align}
where we used $\lambda_h\simeq m_h^2/2v_h^2$ and $\lambda_s\simeq m_s^2/2v_s^2$. The quantity $\Gamma_h(m_s)$ is the Higgs decay width evaluated at the singlet scalar mass $m_s$~\cite{Djouadi:2005gi}. 
For $m_s>2m_t$, the dominant decay channel is into top pairs $\Gamma_h(m_s>2m_t)\simeq (3/8
\pi)m_s(m_t/v_h)^2 \simeq 7~{\rm GeV}~(m_s/m_h)$~\cite{Djouadi:2005gi}. We calculate for $m_s\gg m_h$:
\begin{eqnarray}
\frac{\Gamma_h(m_s)\sin^2(\theta_{hs})}{\Gamma_{s\to hh}} \simeq 30\frac{\Gamma_h(m_s)}{10~{\rm GeV}~(m_s/m_h)}\left(\dfrac{m_h}{ m_s}\right)^{\!2},
\end{eqnarray}
so that the $\Gamma_{s\to hh}$ dominates for $m_s\gg m_h$. Instead, for $m_s< 2m_h$, the decay into Higgs pairs is kinematically forbidden and the singlet scalar $s$ decays through Higgs mixing with the decay width $\Gamma_h(m_s)\sin^2(\theta_{hs})$~\cite{Djouadi:2005gi,DEramo:2024lsk}.
To prevent large quadratic correction to the Higgs mass,
\begin{equation}
\label{eq:Delta_m_H}
    \Delta m_H^2 = \frac{\lambda_{hs}v_s^2}{2}~\lesssim ~(125~\rm GeV)^2 \; ,
\end{equation}
the mixing must be smaller than:
\begin{equation}
\label{eq:naturalness_requirement}
    \lambda_{hs} \lesssim \lambda_{hs}^{\rm max}\simeq 3\times 10^{-16}\left( \frac{10^9~\rm GeV}{v_s} \right)^2  .
\end{equation}
In turn, it leads to a very long lifetime for $s$:
\begin{align}
    \frac{\Gamma_{s\to hh}}{H_{\textrm{$s$-dom}}} &\simeq \frac{1 }{\lambda_s\mathcal{A}\,C_d^{1/2\!}\alpha_{\rm ann}^{3/2}}\left( \frac{\lambda_{hs}}{\lambda_{hs}^{\rm max}} \right)^{\!2} \left(\frac{2.0\times 10^{7}~\rm GeV}{v_s}\right)^{\!6} ,\label{eq:Gamma_H_s-dom}
\end{align}
where $H_{\textrm{$s$-dom}}$ is the Hubble expansion rate when the universe becomes dominated by the scalar field $s$, below the temperature $T_{\textrm{$s$-dom}}$, assuming that $s$ is long-lived enough.
Supposing that the energy density of $s$ redshifts like matter in a radiation-dominated universe, then $T_{\textrm{$s$-dom}}$ is related to the DW annihilation temperature by 
\begin{equation}
\label{eq:T_S_dom}
   T_{\textrm{$s$-dom}}\simeq \alpha_{\rm ann}^{\rm DW+V}T_{\rm ann}\;  .
\end{equation}
The naturalness requirement in Eq.~\eqref{eq:naturalness_requirement} implies that for
\begin{equation}
    v_{s}~\gtrsim~ \frac{6\times 10^{7}~\rm GeV}{(\lambda_s\mathcal{A}C_s^{1/2})^{1/6}}\left( \frac{0.01}{\alpha_{\rm ann}}\right)^{1/4},
\end{equation}
the DW annihilation would be followed by a period of matter domination, as shown in Eq.~\eqref{eq:Gamma_H_s-dom}. This would lead to entropy injection, causing a dilution of the GW and PBH abundances~\cite{Gouttenoire:2023pxh}. For $v_{s}\gtrsim 2.5\times 10^{10}~\rm GeV$, the reheating after the matter era would even occur below $T\simeq 1~\rm MeV$. We can write:
\begin{align}
    \frac{\Gamma_{s\to hh}}{H_{\rm BBN}} \simeq \frac{1}{\lambda_s^{1/2}} \left( \frac{\lambda_{hs}}{\lambda_{hs}^{\rm max}} \right)^{\!2} \left(\frac{2.5\times 10^{10}~\rm GeV}{v_s}\right)^{\!3}\; \label{eq:Gamma_H_BBN},
\end{align}
where $H_{\rm BBN}$ is the Hubble expansion rate at the onset of BBN when the temperature is $T\simeq 1~\rm MeV$. For $s$ to fully decay before BBN, one should have $\Gamma_{s\to hh}/H_{\rm BBN} \gtrsim 1$, which leads to a constraint on $v_s$. However, in this paper, we assume for simplicity that the solution to the naturalness puzzle --- see, e.g.\ \cite{Giudice:2013yca,Barbieri:2013vca,Dine:2015xga,Gouttenoire:2022gwi,DAgnolo:2022mem} for reviews --- is not compromised by the large quadratic contribution in Eq.~\eqref{eq:Delta_m_H}, such that the bound in Eq.~\eqref{eq:naturalness_requirement} can be relaxed without introducing a period of matter domination following DW annihilation. Alternatively, it should be possible to keep the naturalness requirement in Eq.~\eqref{eq:naturalness_requirement} without introducing a matter era by adding new dark states into which the singlet $s$ can decay; see for example Ref.~\cite{Baldes:2023rqv}. We refer the reader interested in the impact of a matter era on GW and PBH detectability to Ref.~\cite{Gouttenoire:2023pxh}. The results of Ref.~\cite{Gouttenoire:2023pxh}, realized in the context of first-order phase transitions, should apply straightforwardly to the DW network scenario discussed here.

\subsection{Particle Dark Matter}
\label{sec:darkmatter}
If the lifetime of the singlet scalar in Eq.~\eqref{eq:def_Gamma_s} is much longer than the age of the universe $\Gamma_s^{-1}\gg 13.8~\rm Gyr$, then the scalar $s$ can contribute to DM~\cite{DEramo:2024lsk}.
Assuming that all the energy stored in the DW network is converted into particles $s$, their abundance today -- if stable -- is given by
\begin{equation}
\label{eq:Omega_s_1}
 \Omega_s = \alpha_{\rm ann}^{\rm DW+V}\frac{a(T_{\rm eq})}{a(T_{\rm NR})}  ,
\end{equation}
where $T_{\rm eq}\simeq 0.8~\rm eV$ and  $T_{\rm NR}(k)$ is the temperature when the $s$ particles with momentum $k$ at $T_{\rm ann}$ become non-relativistic and start redshifting like matter. We have 
\begin{equation}
    T_{\rm NR}(k) =  T_{\rm ann}/\gamma_s,
\end{equation}
where $\gamma_s$ is the typical Lorentz factor of particles $s$ with momentum $k$ right after DW annihilation. Lattice simulations~\cite{Chang:2023rll} have calculated the Fourier spectrum of $\rho_s(k)$ and found a non-relativistic component around $\gamma_s\simeq 1$ and a relativistic component around $\gamma_s\sim 4$. The $s$-abundance depends on the relative abundance of those two populations.\footnote{We leave for future studies a more precise derivation of the dark matter abundance using the integration over the density power spectrum $d\rho_s(k)/dk$, leading to $\Omega_s = \alpha_{\rm ann}^{\rm DW+V}\left<\rho_s\right>^{-1}\int dk (d\rho_s(k)/dk) (a(T_{\rm eq})/a(T_{\rm NR}(k))$.} Waiting for future studies we simply takes $\gamma_s\sim 2$, which aligns with the choice made in another study~\cite{DEramo:2024lsk}. 
Eq.~\eqref{eq:Omega_s_1} becomes
\begin{equation}
 \Omega_s = \frac{\alpha_{\rm ann}^{\rm DW+V}}{\gamma_s}\left(\frac{g_{\star,s}(T_{\rm ann})}{g_{\star,s}(T_{\rm eq})}\right)^{\!1/3}\left(\frac{T_{\rm ann}}{T_{\rm eq}}\right)  .
\end{equation}
\begin{table*}[t!]
  \begin{center}
    \begin{tblr}{|Q[c,0.1cm]|Q[c,1.8cm]|Q[c,1.3cm]|Q[c,1.3cm]|Q[c,1.3cm]|Q[c,1.3cm]|Q[c,1.3cm]|Q[c,1.7cm]|Q[c,1.5cm]|Q[c,0.9cm]|}
      \hline
      \SetCell[r=4,c=2]{c}{{{\textbf{Benchmark models}}}}
    && \SetCell[r=2,c=2]{c}{{{\textbf{QG parameters}}}}&& 
\SetCell[r=2,c=3]{c}{{{\textbf{DW parameters}}}}&& &
\SetCell[r=2,c=3]{c}{{{\textbf{PBH parameters}}}}&  \\\hline
&&&&&&&\\
        \hline
         &&  \SetCell[r=2]{c}{{{\textrm{QG scale }$\Lambda_{\rm QG}$\textrm{ [GeV]}}}}&\SetCell[r=2]{c}{{{\textrm{Scalar VEV }$v_s$\textrm{ [GeV]}}}}&
\SetCell[r=2]{c}{{{\textrm{Surface tension }$\sigma^{1/3}$\textrm{ [GeV]}}}}& 
\SetCell[r=2]{c}{{{\textrm{Bias energy density }$V_{\rm bias}^{1/4}$\textrm{ [GeV]}}}}& \SetCell[r=2]{c}{{{\textrm{DW energy fraction }$\alpha_{\rm ann}$}}}&
\SetCell[r=2]{c}{{{\textrm{Mass }$M_{\rm PBH}$\textrm{ [$M_{\odot}$]}}}}& 
\SetCell[r=1,c=2]{c}{{{\textrm{Abundance}}}} \\\hline
&&&&&&&&\textrm{At formation:} $\beta_{\rm PBH}$&\textrm{Today:} $f_{\rm PBH}$ \\
        \hline
       1& NG15 fit \vspace{-0.3cm}& $1.6\times 10^{30}$ & $3.2\times 10^{5}$ & $3.1\times 10^{5}$& $0.21$& $0.19$ &  $3.2$& $3.4\times 10^{-10}$ &0.06  \\
        \hline
        \SetCell[r=2]{c}{{2}}& \SetCell[r=2]{c}{{PBH DM (heaviest)}} & \SetCell[r=2]{c}{{$4.0\times 10^{26}$}} & \SetCell[r=2]{c}{{$1.0\times 10^{9}$}} & \SetCell[r=2]{c}{{$9.8\times 10^{8}$}}& \SetCell[r=2]{c}{{$4.0\times 10^{4}$}}& \SetCell[r=2]{c}{{$0.15$}} & \SetCell[r=2]{c}{{$8.9\times 10^{-11}$}}& \SetCell[r=2]{c}{{$6.9\times 10^{-14}$}}& \SetCell[r=2]{c}{{$1.0$}}  \\
         &&&&&&&&\\
        \hline
        \SetCell[r=2]{c}{{3}}& \SetCell[r=2]{c}{{PBH DM (lightest)}} & \SetCell[r=2]{c}{{$3.5\times 10^{24}$}} & \SetCell[r=2]{c}{{$1.0\times 10^{11}$}}& \SetCell[r=2]{c}{{$9.8\times 10^{10}$}}& \SetCell[r=2]{c}{{$4.1\times 10^{7}$}} & \SetCell[r=2]{c}{{$0.13$}} &\SetCell[r=2]{c}{{$8.2\times 10^{-17}$}} & \SetCell[r=2]{c}{{$4.6\times 10^{-17}$}} & \SetCell[r=2]{c}{{$1.0$}}   \\
         &&&&&&&&\\
        \hline
    \end{tblr}
    \caption{\label{tab:table_values} QG parameters $\Lambda_{\rm QG}$ and $v_s$, together with the corresponding DW and PBH parameters of the three benchmark models presented in Figs.~\ref{fig:fit-DW2}-\ref{fig:BrokenZ2_QG_PBHs}. We fixed $\lambda_s=1$.}
  \end{center}
\end{table*}
The correct DM abundance $ \Omega_s\simeq 0.26$ is produced for
\begin{equation}
    \alpha_{\rm ann}^{\rm DW+V} \simeq 1.6\times 10^{-10}\left(\frac{\gamma_s}{2}\right) \!\left(\frac{70}{g_{\star,s}}\right)^{\!1/3}\!\!   \left(\frac{1~\rm GeV}{T_{\rm ann}}\right),\!\!\!\!
\end{equation}
where the energy fraction in the DW network accounting for the surface + volume energy $\alpha_{\rm ann}^{\rm DW+V}$ is related to the energy fraction accounting for the surface energy only $\alpha_{\rm ann}\equiv \alpha_{\rm ann}^{\rm DW}$ through $\alpha_{\rm ann}^{\rm DW+V}\simeq 5\alpha_{\rm ann}/2$, cf. Eq.~\eqref{eq:alpha_ann_DW_V}. 
In contrast to singlet scalar DM with vanishing VEV $\left<S\right> = 0$~\cite{McDonald:1993ex,Burgess:2000yq}, the finite VEV $\left<S\right>\neq 0$ allows DM to decay which is constrained by X-rays, CMB, $\gamma$-ray and radio observations~\cite{DEramo:2024lsk}. We assume the Higgs mixing $\theta_{hs}$ in Eq.~\eqref{eq:theta_hs} to be sufficiently small to avoid DM decay constraints and thermal DM production~\cite{DEramo:2024lsk}. Within such assumption $\theta_{hs}\ll 1$, the parameter space where the scalar relic $s$ produced from DW annihilation can satisfy the correct relic abundance is shown with the dashed brown line in Fig.~\ref{fig:BrokenZ2_QG_PBHs}.
Finally, we note that DW networks can also produce baryons and explain the origin of the matter-antimatter asymmetry~\cite{Daido:2015gqa,Mariotti:2024eoh,Vanvlasselaer:2024vmi,Azzola:2024pzq}.

\subsection{Gravitational Waves}\label{sec:GW formation}

\begin{figure*}[t!]
  \centering
  \includegraphics[width=0.85\linewidth]{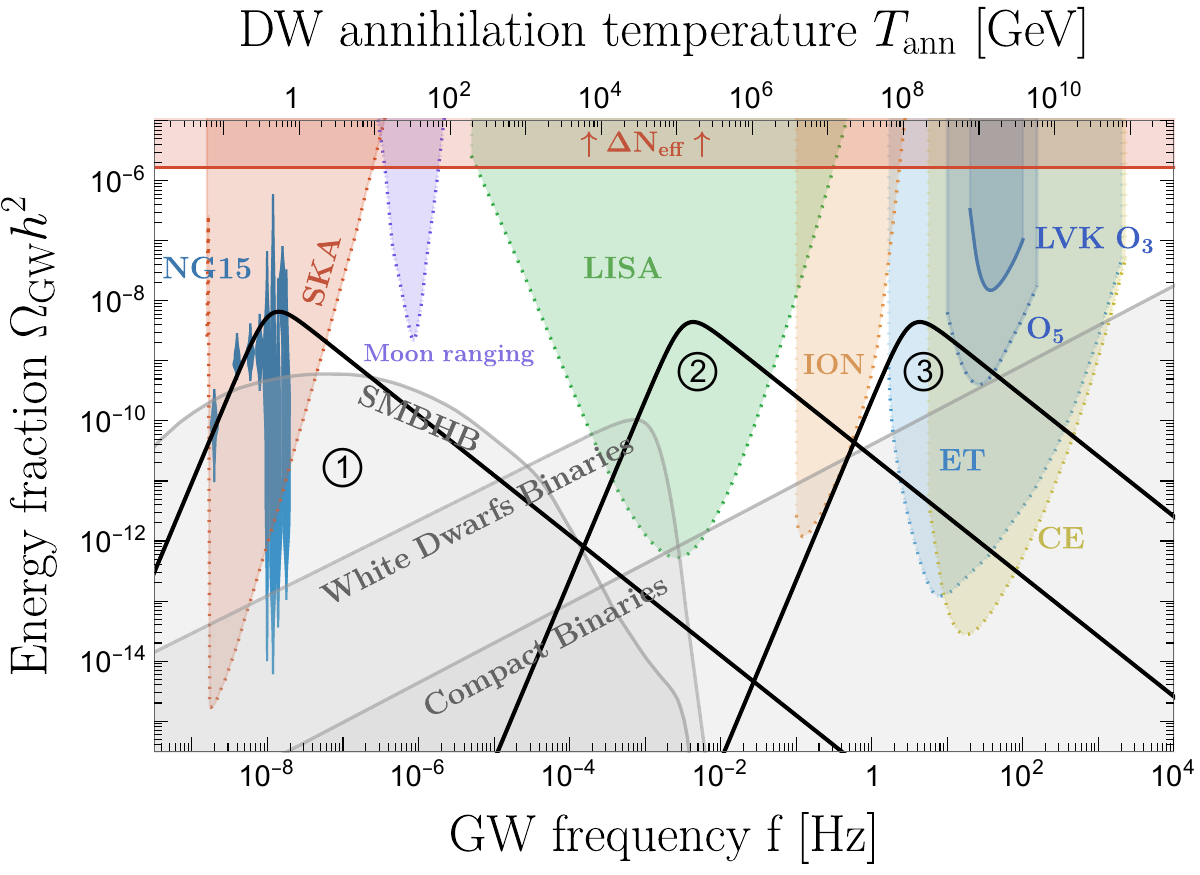}
\caption{The GW spectra in Eq.~\eqref{eq:GW_DW_ann} originating from the annihilation of domain walls together with the sensitivities reached by the different on-going (\textbf{solid}) and planned (\textbf{dotted}) GW experiments and the expected astrophysical GW foregrounds (see text for details). Model 1 can explain the recently detected NG15 signal~\cite{NANOGrav:2023gor,Gouttenoire:2023ftk}, while models 2 and 3 can explain 100$\%$ of DM in terms of PBHs produced from DW annihilation, see Tab.~\ref{tab:table_values} for more details. The three benchmark models are also indicated with stars $\star$ in Fig.~\ref{fig:BrokenZ2_QG_PBHs}.}
  \label{fig:fit-DW2}
\end{figure*}

\paragraph*{GW spectrum.}
During the annihilation process, DWs accelerate to relativistic speeds and emit GWs~\cite{Vilenkin:1981zs, Preskill:1991kd, Chang:1998tb, Gleiser:1998na, Hiramatsu:2012sc, Gelmini:2021yzu, Li:2023gil, Kitajima:2023cek, Takahashi:2008mu, Kitajima:2023kzu,Sakharov:2021dim,Ghoshal:2025iyd}. The GW power spectrum $\Omega_{\rm GW}^0$ observed today is linked to the power spectrum at the time of annihilation, $\Omega_{\rm GW}^{\rm ann}$ by:
\begin{equation}
\Omega_{\rm GW}^0 h^2 = \mathcal{D} \, \Omega_{\rm GW}^{\rm ann}  \; , 
\end{equation}
where $\mathcal{D} \equiv \rho_{\rm rad}(T_{\rm ann})/\rho_{\rm rad}(T_0)$ represents the redshift factor of the universe radiation energy density, given by $\rho_{\rm rad}(T) = \pi^2 g_{\star} T^4 / 30$ under the assumption of adiabatic evolution ($T \propto g_{\star, s}^{-1/3} a^{-1}$), namely,
\begin{align}
\mathcal{D} = \Omega_{\rm rad}^0 h^2 \left( \frac{g_{\star}(T_{\rm ann})}{g_{\star}(T_0)} \right) \left( \frac{g_{\star, s}(T_0)}{g_{\star, s}(T_{\rm ann})} \right)^{4/3} .
\end{align}
The radiation energy fraction today is $\Omega_{\rm rad}^0 h^2 = h^2 \rho_{\rm rad}(T_0) / \rho_0 \simeq 4.21 \times 10^{-5}$, where we used the temperature of the current universe $T_0 \simeq 2.73~\rm K$ \cite{ParticleDataGroup:2024cfk}, the current critical density $\rho_0 = 3 M_{\rm pl}^2 H_0^2$ with $H_0 \simeq 100 h~\rm km/s/Mpc$ and $h$ being the reduced Hubble constant, and $g_{\star}(T_0) \simeq 3.38$. Additionally, we have $g_{\star, s}(T_0) \simeq 3.94$, which leads to:
\begin{align}
\mathcal{D} &\simeq 1.69 \times 10^{-5} \left( \frac{g_{\star}(T_{\rm ann})}{106.75} \right) \left( \frac{106.75}{g_{\star, s}(T_{\rm ann})} \right)^{4/3}  .
\end{align}

The GW spectrum produced by long-lived DWs annihilating at $t_{\rm ann}$ follows the quadrupole formula \cite{Hiramatsu:2010yz, Kawasaki:2011vv, Hiramatsu:2013qaa, Saikawa:2017hiv}:
\begin{equation}
\label{eq:GW_DW_ann}
\Omega_{\rm GW}^{\rm ann} = \epsilon_{\mathsmaller{\rm GW}} \frac{G \mathcal{A}^2 \sigma^2}{\rho_{\rm rad}(t_{\rm ann})} S(f) = \frac{3}{32 \pi} \epsilon_{\mathsmaller{\rm GW}} \alpha_{\rm ann}^2 S(f) \; ,
\end{equation}
with $\epsilon_{\mathsmaller{\rm GW}} = 0.7 \pm 0.4$ \cite{Hiramatsu:2013qaa}. The spectral function $S(f)$ should exhibit an infrared slope $\Omega_{\rm GW} \propto f^3$ to comply with causality \cite{Durrer:2003ja, Caprini:2009fx, Cai:2019cdl, Hook:2020phx} and an ultraviolet slope $\Omega_{\rm GW} \propto f^{-1}$ as suggested by lattice simulations \cite{Hiramatsu:2013qaa}. This motivates modeling $S(f)$ with the following smoothing function:
\begin{equation}
S(f) = \frac{2}{(f/f_{\rm peak}) + (f_{\rm peak}/f)^3} \; ,
\end{equation}
where the peak frequency $f_{\rm peak}$ is determined by the Hubble factor at annihilation, redshifted to the present day as~\cite{Hiramatsu:2013qaa}:
\begin{align}
f_{\rm peak} &= \frac{a(t_{\rm ann})}{a(t_0)} H(t_{\rm ann}) \\
&\simeq 1.6~{\rm mHz} \!\left( \frac{g_{\star}(T_{\rm ann})}{106.75} \right)^{\!\!\frac{1}{2}\!\!} \!\left( \frac{106.75}{g_{\star, s}(T_{\rm ann})} \right)^{\!\!\frac{1}{3}\!\!} \!\left( \frac{T_{\rm ann}}{10~\rm TeV} \right)\; .\notag
\end{align}

\begin{figure*}[t!]
  \centering
    \includegraphics[width=0.9\linewidth]  {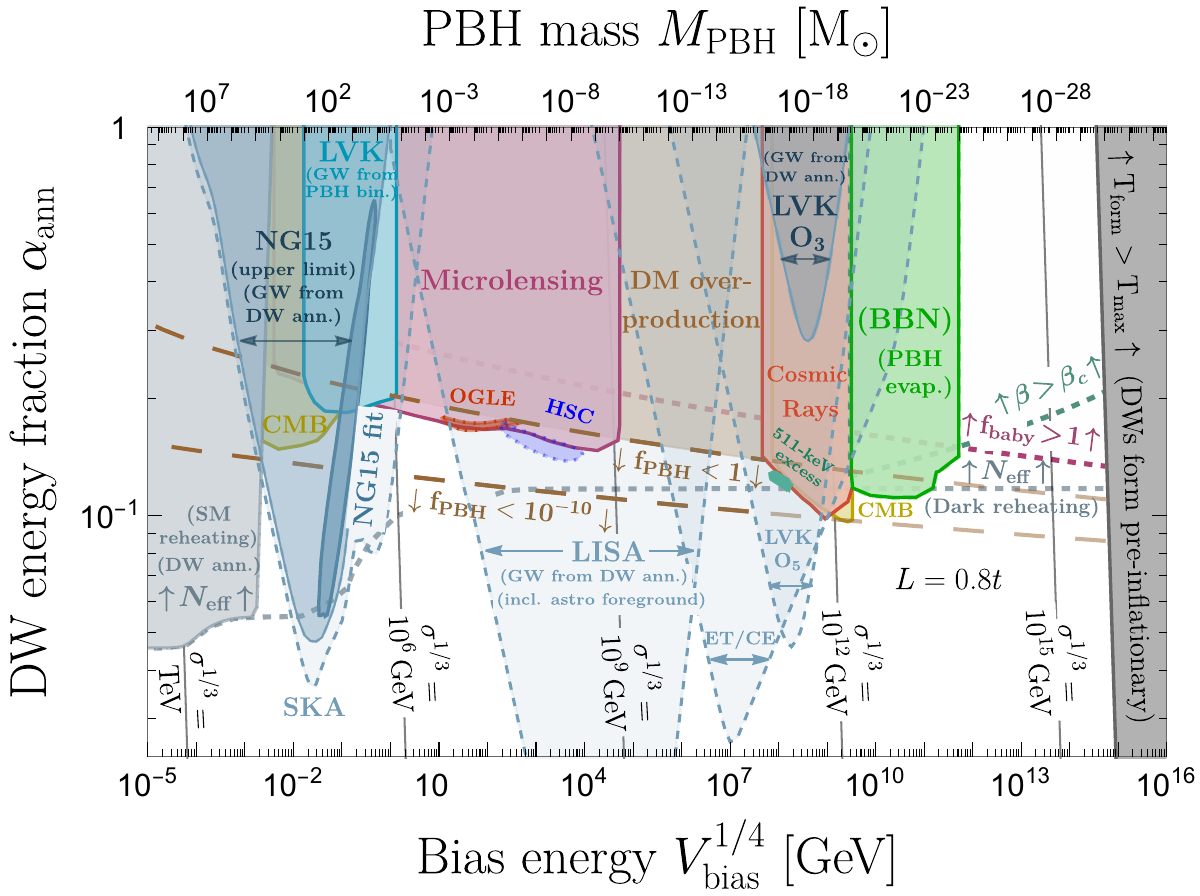}
  \caption{ \textbf{Model-independent} cosmological consequences of a DW network annihilating with energy fraction $\alpha_{\rm ann}=\rho_{\rm DW}/\rho_{\rm rad}$ under the effect of a bias energy density $V_{\rm bias}$. The \textbf{light gray} region shows the BBN constraints in Eq.~\eqref{eq:Delta_N_eff_app}. The \textbf{dashed blue} regions show the GW constraints including the astrophysical foregrounds, which are displayed in Fig.~\ref{fig:fit-DW2}. We show the regions for which PBHs are excluded by CMB observations~\cite{Ali-Haimoud:2016mbv,Poulin:2017bwe,Serpico:2020ehh} (\textbf{yellow}), LIGO-Virgo-Kagra~\cite{LIGOScientific:2019kan,DeLuca:2020qqa} (\textbf{blue}), microlensing constraints  \cite{CalchiNovati:2013jpj,CalchiNovati:2013jpj,Niikura:2017zjd,Smyth:2019whb,Sugiyama:2019dgt} (\textbf{purple}) with the two super-imposed {\bf dotted} regions being the best-fit of the PBH interpretation for HSC and OGLE events \cite{Niikura:2019kqi,Sugiyama:2021xqg,Kusenko:2020pcg}. The {\bf brown} boundary can explain 100$\%$ of DM. In the {\bf red} region, Hawking evaporation leads to an overproduction of cosmic rays \cite{DeRocco:2019fjq,Laha:2019ssq,Keith:2021guq,Carr:2009jm,Boudaud:2018hqb,Laha:2020ivk,Dasgupta:2019cae,Coogan:2020tuf,Korwar:2023kpy}, yet it may also contribute to the $511~\rm keV$ excess indicated in {\bf green} \cite{DeRocco:2019fjq,Laha:2019ssq,Keith:2021guq}. Finally, the lack of observed energy injection from Hawking radiation in the CMB \cite{Poulin:2016anj,Stocker:2018avm,Poulter:2019ooo} and BBN \cite{Carr:2009jm,Acharya:2020jbv,Keith:2020jww} excludes the {\bf yellow} and {\bf green} regions. Above the \textbf{dashed purple} line, at least one baby universe is produced in our past lightcone $f_{\rm baby}>1$ leading to the possibility of a multiverse~\cite{Linde:2015edk}. Above the \textbf{dashed green} line, PBHs dominate the energy density of the universe before evaporating $\beta>\beta_c$. We chose $\ell \equiv L/t = 0.8$ in Eq.~\eqref{eq:F_function_PBH}, a conservative value in terms of PBH production. In the \textbf{dark gray} region on the right, the DW formation temperature is larger than the maximal temperature allowed by BICEP/Keck~\cite{BICEP:2021xfz} $T_{\rm form}>T_{\rm max}$, see Eq.~\eqref{eq:T_max}.
 These constraints are applied to the $\mathbb{Z}_2$ singlet with Quantum Gravity bias in Fig.~\ref{fig:BrokenZ2_QG_PBHs}. }
  \label{fig:DW_generic}
\end{figure*}

\begin{figure*}[t!]
  \centering 
\includegraphics[width=0.9\linewidth]{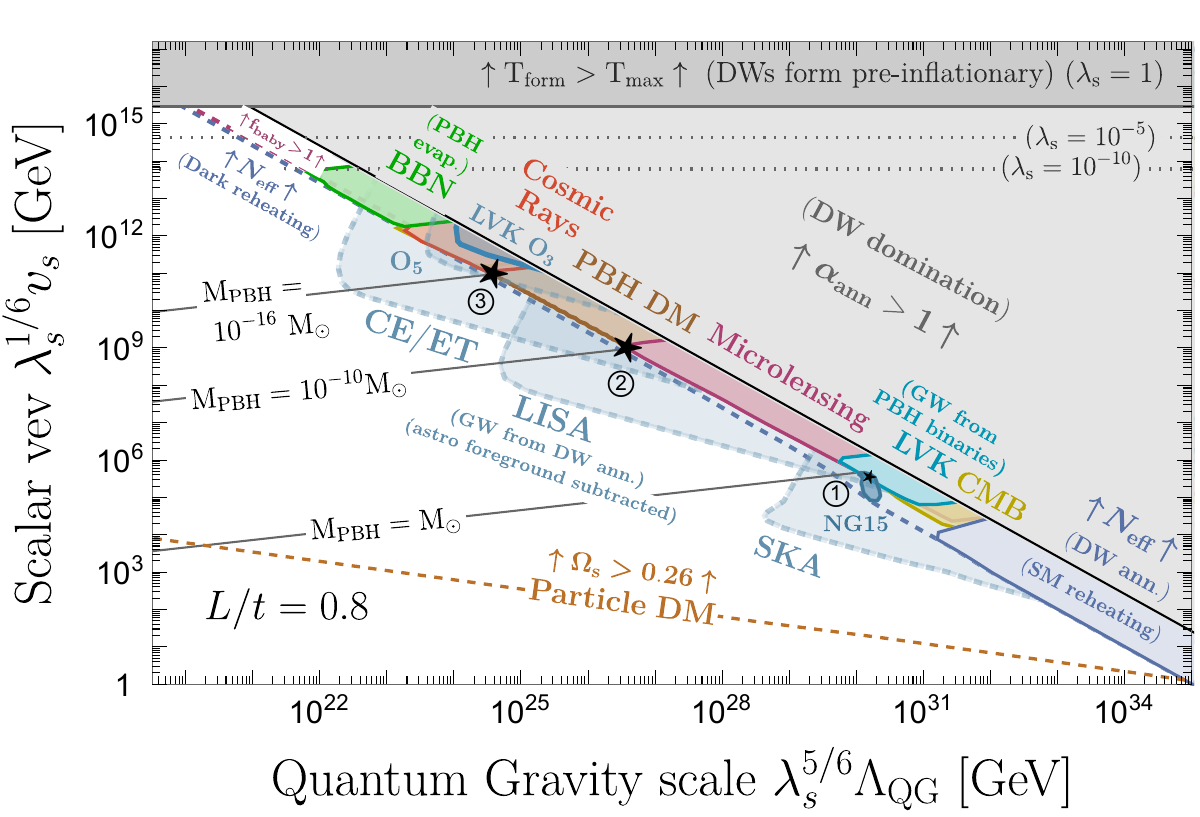}  
  \caption{Application of the $N_{\rm eff}$, GW, PBH constraints and baby universe efficient production regions of Fig.~\ref{fig:DW_generic} to the parameter space of the $\mathbb{Z}_2$-symmetric singlet broken by QG, cf. Eq.~(\ref{eq:combination}). In the \textbf{top-right gray} triangle, DW networks annihilate after dominating the universe, which would lead to a universe very different from ours. On the \textbf{boundary} of it, DW networks annihilate with a significant energy density fraction, leading to efficient particles, GWs, PBHs and baby universe production. In contrast to Fig.~\ref{fig:DW_generic} where the constraints on GW from DW annihilation assume that the astrophysical foreground -- shown in gray in Fig.~\ref{fig:fit-DW2} -- can not be subtracted (pessimistic approach), the GW constraints of the present plot assume complete astrophysical foreground subtraction (optimistic approach). Along the dashed \textbf{orange} lines labeled ``Particle DM'' the scalar particles $s$ can be produced from DW annihilation with the observed relic abundance, see Sec.~\ref{sec:darkmatter}. }
  \label{fig:BrokenZ2_QG_PBHs}
\thisfloatpagestyle{empty} 
\end{figure*}

\paragraph*{GW detection prospects.}
The GW spectra produced from DW annihilation are shown in Fig.~\ref{fig:fit-DW2} for three benchmark points, for which the parameter choices are listed in Tab.~\ref{tab:table_values}. We display the different regions excluded by ground-based interferometers LIGO-Virgo-KAGRA (LVK) run O3 \cite{KAGRA:2021kbb}, detection potential for future LVK run O5~\cite{LIGOScientific:2014pky}, future pulsar timing array SKA \cite{Janssen:2014dka}, the proposed Moon ranging mission~\cite{Blas:2021mqw,Blas:2021mpc}, prospects from atom interferometer AION km~\cite{Badurina:2019hst,Proceedings:2023mkp}, upcoming space-based interferometers LISA \cite{Audley:2017drz, Robson:2018ifk}, ET \cite{Punturo:2010zz, Maggiore:2019uih}, and CE \cite{Reitze:2019iox}. The BBN bound is derived from $\Delta N_{\rm eff}\lesssim 0.34$ following App.~A in \cite{Gouttenoire:2019kij}.  The Power-Law Integrated Curves (PLIC) are built assuming the signal-to-noise ratios (SNR) and observation times $T$ of $\rm ( SNR = 2,~\mathit{T}=160~days)$ for LVK O3,  $\rm (SNR=10,~\mathit{T}=1 ~years)$ for LVK O5, $\rm (SNR=2,~\mathit{T}=15 ~years)$ for Moon ranging, $\rm (SNR=10,~ \mathit{T}=5 ~years)$ for AION km, $\rm (SNR=10,~\mathit{T}=10 ~years)$ for SKA, LISA, ET, and CE. 
 We also show the expected foregrounds from galactic white dwarf binaries modeled using the modeling in \cite{Robson:2018ifk} (see  \cite{Lamberts:2019nyk, Boileau:2021gbr} for different modeling), extragalactic supermassive black hole binaries (SMBHB) taken from \cite{Rosado:2011kv}, along with extragalactic compact binaries (neutron stars and black holes) modelled as a power-law $\Omega \propto f^{2/3}$ fitted on LIGO O3 data \cite{KAGRA:2021kbb}. 
We show two types of GW constraints, according to whether the astrophysical foregrounds are irreducible, see Fig.~\ref{fig:DW_generic}, or can be fully subtracted, see Fig.~\ref{fig:BrokenZ2_QG_PBHs}. A region is excluded if the associated GW spectrum in Eq.~\eqref{eq:GW_DW_ann} exceeds the sum of the corresponding PLIC and the eventual astrophysical foregrounds. In Fig.~\ref{fig:fit-DW2}, we also show the violin posteriors of the GW signal found in NANOGrav 15-year (NG15) data in June 2023~\cite{NANOGrav:2023gor, Antoniadis:2023rey, Reardon:2023gzh, Xu:2023wog, InternationalPulsarTimingArray:2023mzf}. The best fit of the DW interpretation~\cite{ Ferreira:2022zzo,Gouttenoire:2023ftk} of the NG15 signal is shown in Figs.~\ref{fig:DW_generic} and \ref{fig:BrokenZ2_QG_PBHs}.
The two-dip feature in the SKA constraint shown in Fig.~\ref{fig:DW_generic} may seem puzzling. In fact, the higher-frequency peak arises from a sudden change in concavity between the SMBHB foreground and the SKA sensitivity curve, depicted in gray and red, respectively, in Fig.~\ref{fig:fit-DW2}. 

\subsection{Primordial Black Holes}\label{sec:PBH formation}

\paragraph*{Collapse of late-annihilators.}
We now review the calculation of the PBH abundance from DW networks following \cite{Gouttenoire:2023gbn,Gouttenoire:2023ftk}. We model closed DW configuration as spherical false vacuum bubbles containing an energy density given by the potential bias $V_{\rm bias}$. The total mass-energy $M$ of these configurations is given by:
\begin{equation}
\label{eq:M_decomposition}
    M = 4\pi V_{\rm bias} R^3/3+4\pi \sigma \gamma R^2 -8\pi^2 G \sigma^2 R^3 \; ,
\end{equation}
where $G \equiv 1/(8\pi M^2_{\rm pl})$ denotes the gravitational constant, $\gamma$ represents the Lorentz factor, and the three terms in the right-hand side correspond to the contributions from the volume, the surface, and the gravitational binding energy, respectively.
PBHs form when closed DWs shrink within their Schwarzschild radius, at a time $t_{\rm PBH}$ defined by:
\begin{equation}
\label{eq:schwarszchild_radius}
    R_{\rm sch}(t_{\rm PBH}) = 2\, GM(t_{\rm PBH}) \; .
\end{equation}
As in \cite{Gouttenoire:2023gbn,Gouttenoire:2023ftk}, we only consider PBH formation from \textit{late-annihilators}.\footnote{The importance of super-horizon closed DWs, referred to as \textit{late birds}, was first discussed in \cite{Ferrer:2018uiu,Pujolas:2022qvs}.} These are closed DWs with super-horizon sizes at their formation, whose motion is frozen in comoving coordinates, $R(t)\propto a(t)$. Their annihilation is delayed until they finally penetrate the horizon. This delay can allow their Schwarzschild radius, $R_{\rm sch}\propto R(t)^3$, to grow beyond their actual radius $R(t)$. In such cases, they collapse into PBHs as soon as they enter the horizon when $R(t_{\rm PBH}) = t_{\rm PBH}$. We leave for future work the study of PBH formation from the collapse of DWs that are so close-to-spherical that they can shrink significantly and enter their Schwarzschild radius well within the Hubble horizon $R_{\rm sch}(t)< t$, see, e.g. \cite{Ferrer:2018uiu,Dunsky:2024zdo}.
 From the calculation of DW trajectory within the thin-shell approximation in General Relativity, it can be shown \cite{Gouttenoire:2023gbn,Gouttenoire:2023ftk} that the three mass fractions in Eq.~\eqref{eq:M_decomposition} take respectively the values $25\%$, $50\%$ and $25\%$ at $t_{\rm PBH}$, almost independent of the choice of DW model parameters. This implies that we approximately have $M \simeq 16\pi V_{\rm bias} R^3/3$. Plugging this equation into Eq.~\eqref{eq:schwarszchild_radius} together with $R(t_{\rm PBH}) = t_{\rm PBH}$, we obtain that closed DWs collapse into PBHs if they enter the horizon after the time:
\begin{equation}
\label{eq:t_PBH_t_dom}
    t_{\rm PBH} \simeq \frac{\sqrt{3}M_{\rm pl}}{2\sqrt{V_{\rm bias}}}\simeq t_{\rm dom} \;,
\end{equation}
coinciding with the time of false vacuum domination, given by the second equality in Eq.~\eqref{eq:t_dom}. Any DW --- assumed to be spherical --- entering the horizon after $t_{\rm PBH}$ in Eq.~\eqref{eq:t_PBH_t_dom}, unavoidably collapses into PBH.
Tracing back along the DW trajectory from the time of collapse $t_{\rm PBH}$ to the onset of annihilation $t_{\rm ann}$, we obtain the minimal (super-horizon) radius at $t_{\rm ann}$ required for the DW network to collapse into PBHs as:
\begin{equation}
\label{eq:R_PBH_ann}
     \frac{R_{\rm ann}^{\rm PBH}}{t_{\rm ann}} \simeq 0.780\, \log_{10}^2(\alpha_{\rm ann}) -0.618\, \log_{10}(\alpha_{\rm ann}) + 0.407 \; .
\end{equation}
For $t\lesssim t_{\rm ann}$, the effect of the potential bias $V_{\rm bias}$ can be neglected, and the DW network can be well described by the scaling regime in which the network correlation length is given by $L\simeq t$. The fraction of closed DWs --- of any shapes, not necessarily spherical --- which are large enough to contain a spherical false vacuum ball of size $r\equiv R/L$ can be derived from percolation theory~\cite{Gouttenoire:2023gbn,Gouttenoire:2023ftk}:
\begin{equation}
\label{eq:F_r}
    \mathcal{F}(r) \simeq s_{\rm ball}(r) \times p^{s_{\rm ball}(r)} \;,
\end{equation}
where the probability for a lattice site to be occupied in a $\mathbb{Z}_2$-symmetric potential is $p = 0.5$, and the number $s_{\rm ball}(r)$ of lattice sites contained in a ball of size $r$ is~\cite{Gouttenoire:2023gbn,Gouttenoire:2023ftk}:
\begin{equation}
    \frac{s_{\rm ball}(r)}{r^3} \equiv 8+\left(\frac{1+\tanh{(a_1\log{(r/a_2)})}}{2}\right)\left(\frac{4\pi}{3}-8 \right)\;,
\end{equation}
with $a_1 \simeq 1.15$ and $a_2 \simeq 5.55$. A fit of the fraction of closed DWs in Eq.~\eqref{eq:F_r} larger than the critical radius $r_{\rm ann}^{\rm PBH}\equiv R_{\rm ann}^{\rm PBH}/L$ in Eq.~\eqref{eq:R_PBH_ann} is given by~\cite{Gouttenoire:2023gbn,Gouttenoire:2023ftk}:
\begin{equation}
\label{eq:F_function_PBH}
    \mathcal{F}(r_{\rm ann}^{\rm PBH}) \simeq \exp\left[- \frac{a}{\ell^b}\left(\frac{1}{\alpha_{\rm ann}}\right)^{{c}/{\ell^{d}}}\right]\;,
\end{equation}
with $\ell\equiv L/t$ and parameters $a \simeq 0.659$, $b \simeq 2.49$, $c \simeq 1.61$, and $d \simeq 0.195$. 
The normalized density of PBHs relative to the observed dark matter density is:
\begin{equation}
\label{eq:f_PBH_formula}
    f_{\rm PBH} \simeq \left(\frac{T_{\rm dom}}{1~\rm TeV}\right) \frac{\mathcal{F}(r_{\rm ann}^{\rm PBH})}{8 \times 10^{-13}} \;,
\end{equation}
where the temperature at which DWs dominate the energy density of the universe can be expressed as $T_{\rm dom} \simeq 1.23 (M_{\rm pl}/t_{\rm dom})^{1/2}/g_\ast(T_{\rm dom})^{1/4}$. Note that $f_{\rm PBH}$ depends very sensitively on the model parameters. 
The PBH mass is supposed to be given by the DW mass-energy at horizon crossing ($R=2GM$):
\begin{equation}
\label{eq:M_PBH_peak}
    M_{\rm PBH} \simeq \frac{t_{\rm PBH}}{2G} \simeq 14 M_{\odot} \left( \frac{100~\rm MeV}{V_{\rm bias}^{1/4}} \right)^2 \;,
\end{equation}
where $M_{\odot} = 2 \times 10^{30}~{\rm g}$ denotes the solar mass. The above equation indicates that $M_{\rm PBH}$ depends only on $V_{\rm bias}$ but not on $\sigma$. The universal scaling behavior of gravitational collapse \cite{Choptuik:1992jv,Niemeyer:1997mt,Niemeyer:1999ak,Musco:2012au,Musco:2018rwt} suggests that the PBH mass in Eq.~\eqref{eq:M_PBH_peak} is only the peak $M_{\rm PBH}^{\rm peak}$ (up to $\mathcal{O}(1)$ factors to be determined) of an extended mass distribution. A more realistic PBH mass spectrum decreases as a power-law $f_{\rm PBH}\propto M_{\rm PBH}^{1+1/\gamma_c}$ with $\gamma_{c}\simeq 0.36$ at lower masses $M_{\rm PBH}\lesssim M_{\rm PBH}^{\rm peak}$, and is exponentially suppressed at larger masses $M_{\rm PBH}\gtrsim M_{\rm PBH}^{\rm peak}$ (see e.g. \cite{Baldes:2023rqv,Franciolini:2021nvv}). We leave the determination of the PBH mass distribution for future work. The PBH abundance in Eq.~\eqref{eq:f_PBH_formula} is very sensitive to the correlation length $L$. A larger correlation length to Hubble horizon ratio $L/t$ yields a larger PBH abundance. Following Ref.~\cite{Gouttenoire:2023ftk}, we set $L/t \simeq 0.8$, which is a conservative choice with respect to PBH production.

\paragraph*{PBH detection prospects.}

In Fig.~\ref{fig:DW_generic}, we recast all the existing PBH constraints \cite{Carr:2020gox} from observations of BBN, CMB, GWs, cosmic rays, microlensing in the generic plane $(V_{\rm bias}^{1/4}, \alpha_{\rm ann})$ of biased DW network models. We can see that biased DW networks annihilating with an energy fraction $\alpha_{\rm ann}\gtrsim 0.1$ produce PBHs with observable abundance.\footnote{More specifically, the general guideline for producing GWs and PBHs in quantities detectable by future observations is that DW networks should annihilate with an energy fraction $\alpha_{\rm ann}\gtrsim 0.01$ and $\alpha_{\rm ann}\gtrsim 0.1$ respectively \cite{Gouttenoire:2023gbn}. Successful subtraction of astrophysical foregrounds could enhance the sensitivity of GW observatories to  $\alpha_{\rm ann}\gtrsim 0.001$~\cite{Gouttenoire:2023gbn}.} 
Then in Fig.~\ref{fig:BrokenZ2_QG_PBHs}, we apply the previous generic constraints to the specific particle physics model presented in Sec.~\ref{sec:simplified} --- a $\mathbb{Z}_2$-symmetric singlet scalar with QG $\mathbb{Z}_2$-breaking corrections. The model has three parameters, namely, the scalar quartic $\lambda_s$, the scalar VEV $v_s$, and the QG scale $\Lambda_{\rm QG}$. However, as we have mentioned before, the DW parameters $(V_{\rm bias}^{1/4}, \alpha_{\rm ann})$ can be expressed solely in terms of the two independent combinations $\lambda_s^{1/6} v_s$ and $\lambda_s^{5/6}\Lambda_{\rm QG}$ in Eq.~\eqref{eq:combination}. The parameter space leading to $N_{\rm eff}$, GWs and PBHs signatures is indicated in Fig.~\ref{fig:BrokenZ2_QG_PBHs} where the black contour lines denote different PBH masses.

The mass of the PBHs produced depends on the bias term $V_{\rm bias}$. By adjusting the value of $V_{\rm bias}\in [\rm 10^{-3}, ~10^{15}]~\rm GeV$, the mass of PBHs with substantial abundance --- say $f_{\rm PBH}> 10^{-10}$ --- can vary over a wide range $M_{\rm PBH}\in [10^{-30}, 10^{5}]{M_\odot}$, with $10^{-30}M_\odot\simeq 2~\rm kg$. The production of PBHs heavier than $10^{5}M_{\odot}$ is excluded by the BBN bound in Eq.~\eqref{eq:Delta_N_eff_app}, illustrated by the gray region on the left side of Fig.~\ref{fig:DW_generic}. Solar mass PBHs can be correlated to the nHz GW signal detected in PTA~\cite{Gouttenoire:2023ftk}. PBHs lighter than $10^{-30}M_{\odot}$ are excluded by the condition that the temperature $T_{\rm form}$ at which the DW network forms can not be higher than the maximal temperature $T_{\rm max}$ of the universe, illustrated by the gray region on the right side of Fig.~\ref{fig:DW_generic} and top side of Fig.~\ref{fig:BrokenZ2_QG_PBHs}. We fix the formation temperature to $T_{\rm form}\simeq \sigma^{1/3}$ for the model-independent plot in Fig.~\ref{fig:DW_generic},\footnote{The expression $T_{\rm form} \sim \sigma^{1/3}$ is generally expected for models with a single energy scale, where $T_{\rm form} \sim v_s \sim \sigma^{1/3}$. However, this relationship does not hold for axion models in which the axion mass $m_a$ at zero temperature is much smaller than the axion decay constant $f_a$. In such cases, $T_{\rm form} \simeq f_a$ and $\sigma \simeq 9m_a f_a^2$~\cite{Sikivie:1982qv,Vilenkin:1982ks,GrillidiCortona:2015jxo}.} whereas we use $T_{\rm form}\simeq  2v_s$ for the ${\cal Z}_2$-symmetric DW in Fig.~\ref{fig:BrokenZ2_QG_PBHs}. One illustrates the regions where the DW network does not form for the values $\lambda_s = 1$, $10^{-5}$, and $10^{-10}$.

\paragraph*{PBH as Dark Matter.} \label{sec:QGPBH}
Measurements of the CMB anisotropies by Planck satellite not only provide strong confirmation of the existence of DM but also offer the more precise measurement of its abundance today $\Omega_{\rm DM}\simeq 0.1200(12)/h^2\simeq 0.265(7)$~\cite{Aghanim:2018eyx}. There have been many proposals of DM candidates in the literature and among them PBHs have received considerable attention, see, e.g.~\cite{Carr:2016drx, Carr:2020xqk,Carr:2020gox, Green:2020jor}. 
Astrophysical and cosmological constraints have excluded the possibility of PBHs accounting for 100$\%$ of DM, except within the so-called asteroid mass window, $M_{\rm PBH}\in [10^{-16},~10^{-10}]~M_{\odot}=[2\times 10^{17},~2\times 10^{23}]~{\rm g}$~\cite{Carr:2016drx, Carr:2020xqk,Carr:2020gox, Green:2020jor}. The region where PBHs could account for 100$\%$ of DM is highlighted in brown in Figs.~\ref{fig:DW_generic} and \ref{fig:BrokenZ2_QG_PBHs}.
In terms of the parameters of the $\mathbb{Z}_2$ scalar with QG bias, all the DM can be explained by PBHs for the scalar VEV $v_s\in [10^{9},~10^{11}]~{\rm GeV}$ and the QG scale $\Lambda_{\rm QG}\in [3\times 10^{24},~3\times 10^{26}]~\rm GeV$ where we fixed $\lambda_s = 1$.
 The entire PBH DM region will be probed by LISA, ET/CE, and LIGO O5, see Fig.~\ref{fig:BrokenZ2_QG_PBHs}, even if astrophysical signals can not be subtracted, see Fig.~\ref{fig:DW_generic}.
 
Note, however, that Figs.~\ref{fig:DW_generic} and \ref{fig:BrokenZ2_QG_PBHs} assume instantaneous reheating after DW annihilation. As discussed in Sec.~\ref{sec:matter_era}, the singlet scalars could be feebly coupled to the SM, e.g.\ motivated by the electroweak hierarchy problem, leading DW annihilation to be followed by a matter era. As pointed in Ref.~\cite{Gouttenoire:2023pxh}, this would dilute the GW signal, possibly turning the PBH DM region undetectable by future GW observatories, in contrast to the naive expectation \cite{Cai:2018dig,Bartolo:2018evs}.
It might seem a nightmare scenario but the lighter mass range of the PBH DM region could be explored with future telescopes \cite{e-ASTROGAM:2017pxr,AMEGO:2019gny,Labanti:2021gji} as pointed by \cite{Ray:2021mxu,Ghosh:2021gfa,Keith:2022sow,Malyshev:2022uju,Malyshev:2023oox}, and the presence of the early matter era could be detected using cosmic string archaeology~\cite{Cui:2017ufi,Cui:2018rwi,Gouttenoire:2019rtn,Gouttenoire:2019kij,Gouttenoire:2021wzu,Gouttenoire:2021jhk,Lazarides:2022ezc, Borah:2022vsu,Borah:2023iqo,Ghoshal:2023sfa}.

\begin{figure}[t!]
     \centering
    \includegraphics[width=1\linewidth]{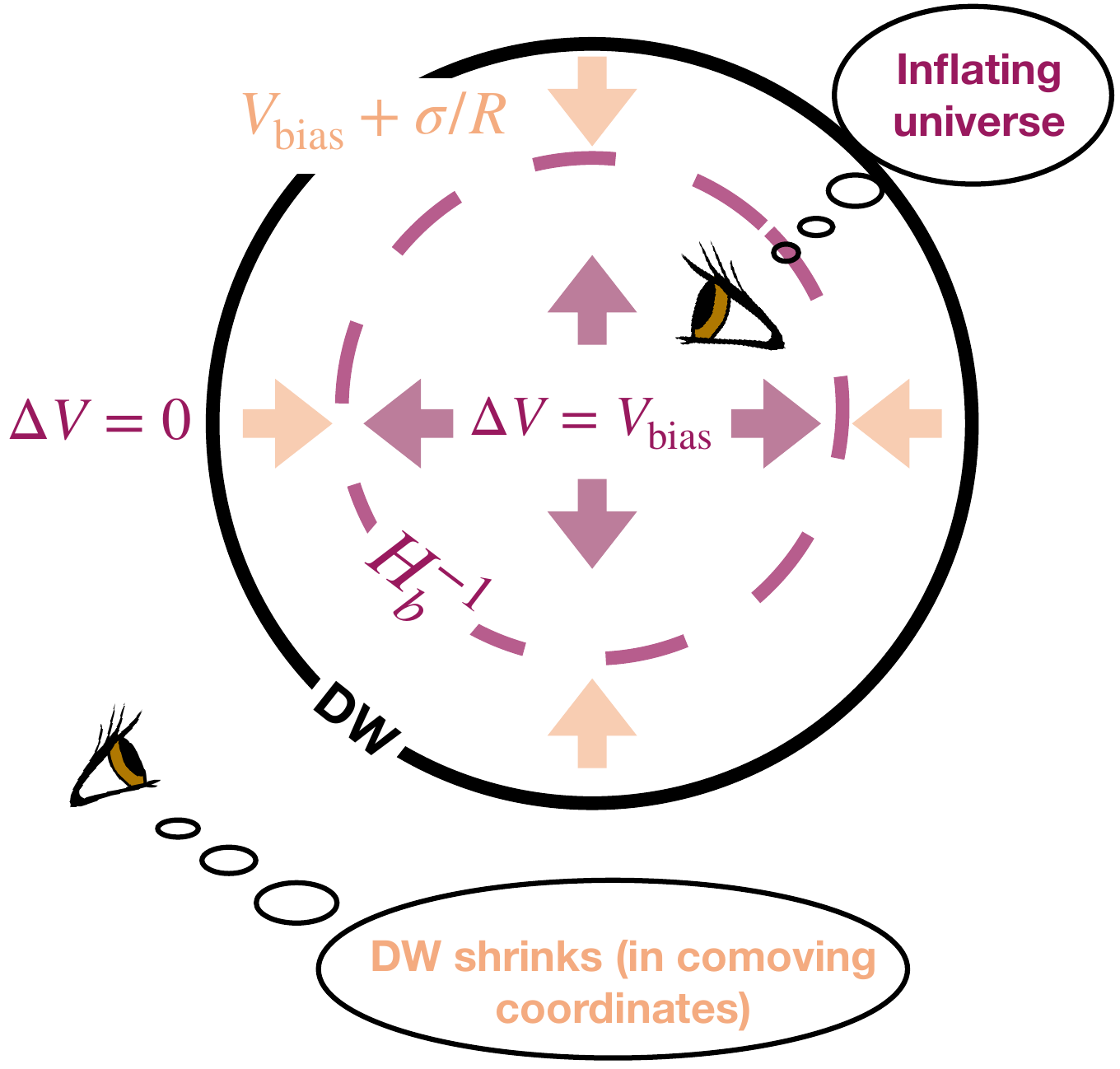}  
     \caption{Production of baby universes by DW networks. This arises when a closed DW configuration -- modeled to be spherical and shown with the \textbf{black} circle -- becomes larger than the Hubble horizon $H_b^{-1}$ of a universe dominated by the bias energy $V_{\rm bias}$ -- shown with the dashed \textbf{purple} circle and given by Eq.~\eqref{eq:H_b}. The \textbf{orange} arrows represent the pressure $V_{\rm bias}+\sigma/R$ driving the DW to locally move inward in the comoving coordinates experienced by an observed located close to the DW. The \textbf{purple} arrows show the volume expansion induced by the bias energy $V_{\rm bias}$ acting as a cosmological constant in the bulk.   }
     \label{fig:slide_wormholes_1} 
 \end{figure}
 \begin{figure*}[t!]
     \centering
    \includegraphics[width=1\linewidth]{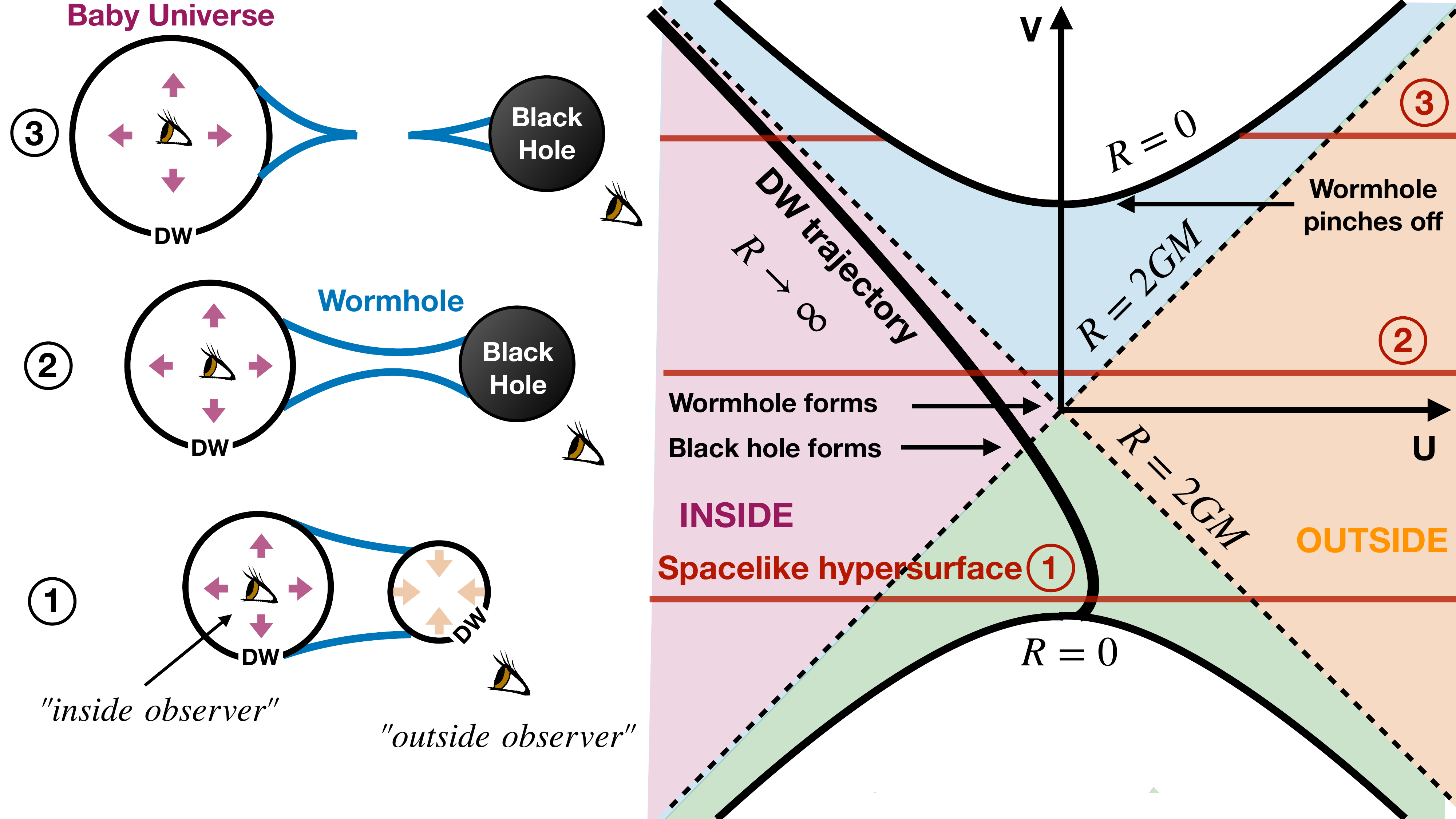}  
     \caption{\textbf{RIGHT}: Diagram representing a Schwarzchild spacetime in Kruskal–Szekeres coordinates (orange+blue) stitched to a de Sitter spacetime in appropriate coordinates (purple+green)~\cite{Blau:1986cw}. The thick black lines illustrate the DW trajectory in such a diagram. While an observer outside the DW perceives the DW collapsing into a black hole, an observer inside the same DW experiences an eternally inflating baby universe. At the initial stage of black hole formation, the baby universe inside the DW remains connected to the parent universe via a wormhole, whose throat corresponds to the black hole interior region shown in \textbf{blue}. An observer inside the wormhole throat would receive signals from both sides before unavoidably ending their trajectory in the singularity at $R=0$.
    Indeed, the wormhole throat pinches off rapidly, ensuring that it is not traversable and that the two observers remain causally disconnected. Consequently, the two observers encountering vastly different spacetime dynamics leads to no paradox.  \textbf{LEFT}: The three drawings on the left side illustrate the three spacelike hypersurfaces associated to the three horizontal red lines in the space-time diagram on the right hand side. }
     \label{fig:slide_wormholes_2} 
 \end{figure*}
\subsection{Baby Universes}\label{sec:wormhole_formation}

Now let us examine DWs that remain larger than the cosmological horizon (i.e. super-horizon) after the time $t_{\rm PBH}$ given in Eq.~\eqref{eq:t_PBH_t_dom}. From the viewpoint of observers outside such a DW, it appears to shrink and eventually collapse into a PBH.
However, for observers inside the DW, the outcome depends on whether the wall radius $R$ is smaller or larger than the cosmological horizon $H_b^{-1}$ in an inflating universe with energy density $V_{\rm bias}$ (the energy density within the DW):
\begin{itemize}
    \item[$\diamond$] If $R < H_b^{-1}$: Observers inside the DW see its boundary moving inward, ultimately forming a black hole around them. They are subsequently drawn into the singularity. 
    \item[$\diamond$] If $R > H_b^{-1}$: The DW boundary lies outside the future light-cone of observers located at its center. They experience an eternally inflating universe with no black holes nor singularity, see Fig.~\ref{fig:slide_wormholes_1}.
\end{itemize}
At first glance, this second scenario seems paradoxical: external observers witness a black hole forming, while internal observers experience an inflating universe. The apparent contradiction is resolved by the geometry of spacetime: as the DW radius expands exponentially, it acts as the boundary of an inflating baby universe connected to the parent universe via a wormhole~\cite{Blau:1986cw,Garriga:2015fdk,Deng:2016vzb}. This wormhole then pinches off, leaving behind the standard black hole singularity in the parent universe and a detached baby universe for those on the inside. The spacetime geometry outside the DW is Schwarzschild with metric
\begin{equation}
    ds^2 = -\left(1-\frac{2GM}{R}\right)dT^2+\left(1-\frac{2GM}{R}\right)^{-1}dR^2+R^2d\Omega^2,
\end{equation}
where $M$ is the mass of the DW. The metric singularity at $R=2GM$ can be removed by introducing the Kruskal-Szekeres coordinates ($U$,$V$,$\theta$,$\phi$)\footnote{The coordinate $V$ should not be confused with the potential $V$ in Eq.~\eqref{eq:lagrangian_tree}.}
\begin{equation}
    U^2-V^2=\left(\frac{R}{2GM}-1 \right)\exp\left( R/2GM\right).
\end{equation}
In Kruskal--Szekeres coordinates, the Schwarzschild spacetime is divided into four distinct regions, see Fig.~\ref{fig:slide_wormholes_2}. The first two, also present in Schwarzschild coordinates, are the black hole exterior $U >0$, $U>|V|$ and interior $V >0$, $V>|U|$. The remaining two, which do not appear in Schwarzschild coordinates, are the white hole region $V < 0$, $|V|>|U|$ and the parallel exterior region $U<0$, $|U|>|V|$. The Schwarzschild horizon is given by $U = \pm V$, while the physical singularity at $R = 0$ is defined by $U^2 - V^2 = -1$.
We have just discussed the spacetime geometry \emph{outside} the DW. \emph{Inside} the DW, however, the vacuum energy bias $V_{\rm bias}$ generates a de Sitter geometry with metric
\begin{equation}
    ds^2 = -(1-H_b^2R^2)dT^2+(1-H_b^2R^2)^{-1}dR^2+R^2d\Omega^2,
\end{equation}
where $H_b$ is the Hubble horizon of this de Sitter universe
\begin{equation}
\label{eq:H_b}
    H_b \equiv \sqrt{\frac{V_{\rm bias}}{3M_{\rm Pl}^2}}.
\end{equation} 
As for the Schwarzschild spacetime, the horizon singularity at $R=H_b^{-1}$ can be removed by introducing the Gibbons-Hawking coordinates ($W$,$X$,$\theta$,$\phi$)
\begin{equation}
    \frac{1-H_bR}{1+H_b R} = W^2-X^2.
\end{equation}
Following~\cite{Blau:1986cw}, we can find another change of coordinate $(W, X)\to  (Y,Z)$ such that the de Sitter half-spacetime and the Schwarzschild half-spacetime can be smoothly stitched together, i.e. $Y=U$ and $Z=V$ along the boundary $U=-V$. We draw the corresponding diagram of the full Schwarzschild-de Sitter spacetime in Fig.~\ref{fig:slide_wormholes_2}. The DW trajectory, shown by the thick black line, illustrates how the DW collapses into a black hole from the perspective of an outside observer, while appearing as the exponentially expanding boundary of an eternally inflating universe for an inside observer.

DWs forming wormholes are the ones that enter the horizon --- assuming spherical DWs --- after the time $t_{\rm baby}$ defined by:
\begin{equation} 
\label{eq:R_baby}
R(t_{\rm baby}) \simeq H_b^{-1} \; ,
\end{equation} 
where $H_b$ is given by Eq.~\eqref{eq:R_baby}.
Solving the DW trajectory in General Relativity~\cite{Gouttenoire:2023gbn}, we can show that the critical time for baby universe formation $t_{\rm baby}$ is related to the critical time for PBH formation $t_{\rm PBH}$ in Eq.~\eqref{eq:t_PBH_t_dom} by: 
\begin{equation} 
t_{\rm baby} \simeq 2t_{\rm PBH} \simeq 2t_{\rm dom}. 
\end{equation} 
Solving the trajectory backward in time, we can show that DW satisfying Eq.~\eqref{eq:R_baby} have a radius at the onset of the annihilation period --- just before the scaling regime becomes violated by the bias pressure --- given by~\cite{Gouttenoire:2023gbn}:
\begin{equation} 
\frac{R_{\rm ann}^{\rm baby}}{t_{\rm ann}} \simeq 0.951~ {\rm \textrm{log}_{10}^2}(\alpha_{\rm ann}) - 0.860~{\rm \textrm{log}_{10}}(\alpha_{\rm ann}) + 1.15. 
\end{equation} 
The critical radius for wormhole formation, $R_{\rm ann}^{\rm baby}$, is $\sim 30\%$ larger than the critical radius for simple PBH formation $R_{\rm ann}^{\rm PBH}$ in Eq.~\eqref{eq:R_PBH_ann}. This makes wormhole production exponentially rarer than simple PBH production.

The number of baby universes $f_{\rm baby}$ in our past light cone is given by~\cite{Gouttenoire:2023gbn}: 
\begin{equation} 
f_{\rm baby} = \mathcal{N}_{\rm patches}(T_{\rm dom}) \times \mathcal{F}(r_{\rm ann}^{\rm baby}), 
\end{equation} 
with $r_{\rm ann}^{\rm baby}\equiv R_{\rm ann}^{\rm baby}/L(t_{\rm ann})$ and:
\begin{equation} 
\mathcal{N}_{\rm patches} \simeq 4.2 \times 10^{38} \left(\frac{g_{\star}(T_{\rm dom})}{100}\right)^{1/2} \left( \frac{T_{\rm dom}}{100~\rm GeV} \right)^3. 
\end{equation} 
Here, $\mathcal{F}(r_{c})$ --- defined in Eq.~\eqref{eq:F_function_PBH} --- is the DW fraction above some threshold radius $r>r_c$.
The parameter space for producing at least one baby universe in our past lightcone is indicated by the dashed purple line in Figs.~\ref{fig:DW_generic} and \ref{fig:BrokenZ2_QG_PBHs}.

Due to quantum transitions, an eternally-inflating baby universe becomes divided in many different exponentially large domains, realizing all possible types of symmetry breaking, and all possible quantum states allowed by the fundamental theory, e.g. all possible types of string theory  compactifications~\cite{Linde:2015edk}. Hence, the baby universes produced from DW networks, studied in this work, are expected to give rise to a multiverse.

The classical trajectory leading to a baby universe inevitably begins with an initial singularity~\cite{Farhi:1986ty}, denoted by $R=0$ in Fig.~\ref{fig:slide_wormholes_2}. This has motivated studies of quantum processes in which a false-vacuum bubble of critical size nucleates via tunneling and subsequently evolves classically~\cite{Hawking:1981fz,Lee:1987qc,Farhi:1989yr,Fischler:1989se}. In our framework, DWs are produced classically once the temperature drops below the critical value in Eq.~\eqref{eq:T_form}. 
We do not, however, make any specific assumption about the origin of the vacuum energy in the scalar potential. 
It could arise, for instance, from quantum tunneling, in which case no singularity would occur~\cite{Hawking:1981fz,Lee:1987qc,Farhi:1989yr,Fischler:1989se}. 
It could also be assumed to have been present since the big bang singularity. 
In the latter case, the singularity at $R=0$ would be of the same type as that encountered in the primordial inflationary paradigm responsible for the CMB and large-scale structure~\cite{Borde:1993xh,Borde:1994ai,Borde:1996pt,Borde:2001nh}. 
Such a singularity can be avoided if the past-averaged Hubble expansion rate,
\begin{equation}
H_{\text{avg}} \;\equiv\; \frac{1}{\tau} \int_{0}^{\tau} H(\lambda)\, d\lambda ,
\end{equation}
is non-positive, $H_{\text{avg}} \leq 0$~\cite{Borde:2001nh}. 
This condition can be realized, for example, in bouncing cosmologies with null energy condition violation~\cite{Creminelli:2006xe,Rubakov:2006pn,Creminelli:2010ba,Graham:2017hfr}, 
or within quantum gravity frameworks~\cite{Khoury:2001wf,Khoury:2001bz,Finelli:2001sr,Ashtekar:2006rx,Ashtekar:2006wn}. Alternatively, the singularity may be avoided in time-symmetric cosmologies such as the Aguirre--Gratton~\cite{Aguirre:2003ck} or Carroll--Chen scenarios~\cite{Carroll:2004pn}. 

To emphasize, the mechanism of baby-universe production from DW networks studied here does not introduce any additional singularities beyond those inherent to any scalar-field dynamics whose potential contains at least one false vacuum minimum. This statement applies broadly to all frameworks involving cosmological DW networks and first-order phase transitions.

\section{Conclusion}\label{sec:conclusion}

In this paper we have studied the rich cosmological consequences of the explicit discrete global symmetry breaking due to QG. For definiteness, we have focused on the addition of a singlet real scalar $S$ with {\em spontaneously} broken $\mathbb{Z}_2$ global symmetry, leading to the formation of DWs. 
Following the swampland global symmetry conjecture, the $\mathbb{Z}_2$ global symmetry of the scalar potential is \emph{explicitly} broken by higher-dimensional operators arising from QG, hence generating a bias energy $V_{\rm bias}$ between $\left<S\right>=+v_s$ and $\left<S\right>=-v_s$. DWs are driven preferably toward the phase containing the higher vacuum energy leading to complete annihilation of the DW networks. If DWs annihilate while constituting a significant energy fraction of the universe $\alpha_{\rm ann}$, then they can produce long-lived dark particles in large abundance and imprint large perturbations on the spacetime metric.

DWs annihilating into dark sectors must have an energy fraction smaller than $\alpha_{\rm ann}\lesssim 0.1$ in order to not violate the $\Delta N_{\rm eff}$ bound. Some of the produced particles can explain DM if $\alpha_{\rm ann}\gtrsim 10^{-10}\times (1{\rm GeV}/T_{\rm ann})$ or could lead to an early matter-dominated era for $\alpha_{\rm ann}\gtrsim 0.01\times (2\times 10^7~{\rm GeV}/v_s)^{1/4}$.

Stochastic GW backgrounds from DW annihilation are louder than astrophysical foregrounds and observable by current and future GW detectors for $\alpha_{\rm ann}\gtrsim 0.1-0.01$.

PBHs are formed from the collapse of closed DWs entering the horizon after the time $t_{\rm dom}$ of false vacuum domination. The fraction of those {\it late-annihilators} becomes significant when biased DW networks annihilate just before dominating the total energy of the universe, i.e. $t_{\rm dom}/t_{\rm ann}\lesssim \mathcal{O}(\rm a~few)$. Hence, biased DW networks annihilating with a large energy fraction $\alpha_{\rm ann}\gtrsim 0.1$, cf. Eq.~\eqref{eq:alpha_DW}, can produce PBHs with an observable abundance. PBHs with mass smaller than $M_{\rm PBH}\simeq 10^{15}~{\rm g}\simeq 5\times 10^{-19}~M_{\odot}$ evaporate in less than $13.8~\rm Gyr$~\cite{Hawking:1974rv}. Heavier PBHs, with masses in the range $M_{\rm PBH}\in [10^{-16}~M_{\odot},10^{-10}~M_{\odot}]$ evade all the astrophysical and cosmological constraints and therefore they can potentially account for $100\%$ of DM.

Finally we studied the production of classical wormholes connected to baby universes.
This phenomenon occurs when DWs become larger than the cosmological horizon of an inflating universe with energy density given by the bias energy $V_{\rm bias}$.
This regime is accompanied with significant productions of PBHs evaporating much before BBN. We leave for future works the possibility of testing this scenario with GW~\cite{Domenech:2024kmh} and particle productions~\cite{Masina:2020xhk}.
The production of baby universes studied in this work -- which first appeared in \cite{Gouttenoire:2023gbn} -- is classical and completely decoupled from the inflaton. In contrast, conventional baby universe production scenarios are relying on quantum or thermal tunneling during inflation \cite{Goncharov:1987ir,Linde:1990ta,Linde:1991sk,Garriga:2015fdk,Deng:2016vzb}. This offers the possibility to produce a multiverse~\cite{Linde:2015edk} within our past light cone from a simple $\mathbb{Z}_2$ symmetry breaking without relying on any inflaton dynamics.\footnote{We refer to Ref.~\cite{Sato:1981gv,Jinno:2023vnr} for baby universe production in the context of first-order phase transitions.}

Fig.~\ref{fig:DW_generic} summaries the cosmological consequences of DW networks in a model-independent way and Fig.~\ref{fig:BrokenZ2_QG_PBHs} focuses on the particular case where the DW bias term arises from QG effects. 

In conclusion, a seemingly minimal setup -- featuring a real scalar field charged under a global $\mathbb{Z}_2$ symmetry and explicitly broken by quantum gravity -- can yield a wide range of cosmological phenomena. It is remarkable that 
QG effects many orders above the Planck scale can 
have such low-energy observational consequences. 

\section*{Acknowledgments}

The authors thank Edoardo Vitagliano for interesting comments on the manuscript. YG acknowledges support by the Cluster of Excellence ``PRISMA+'' funded by the German Research Foundation (DFG) within the German Excellence Strategy (Project No. 390831469), and by a fellowship awarded by the Azrieli Foundation.
SFK, GW, and RR acknowledge the STFC Consolidated Grant ST/X000583/1. SFK also acknowledges the European Union's Horizon 2020 Research and Innovation programme under the Marie Sklodowska-Curie grant agreement HIDDeN European ITN project (H2020-MSCA-ITN-2019//860881-HIDDeN). XW acknowledges the Royal Society as the funding source of the Newton International Fellowship. MY was supported in part by the JSPS Grant-in-Aid for Scientific Research (No.\ 20H05860, 23K17689, 23K25865), and by JST, Japan (PRESTO Grant No.\ JPMJPR225A, Moonshot R\&D Grant No.\ JPMJMS2061). Finally, the authors would also like to express special thanks to the organizers of the Paris workshop on primordial black holes and gravitational waves at Institut Henri Poincaré, Paris, France, where this project was initiated.

\bibliographystyle{apsrev4-1}
\bibliography{refs}
\end{document}